\documentclass[a4paper,11pt]{article}
\pdfoutput=1

\usepackage{jcappub}

\usepackage[T1]{fontenc}

\usepackage{color,soul}
\usepackage{subcaption}
\usepackage{xcolor}
\usepackage{multirow}
\usepackage{makecell}
\usepackage{hyperref}
\usepackage{pifont}
\usepackage{fontawesome}
\usepackage[dvipsnames]{xcolor}
\usepackage{aas_macros}

\title{Constraining Primordial Magnetic Fields with Weak Lensing}

\author[a,b,c]{Anna D'Ambrosio}
\author[a,b,c,d,e]{and Matteo Viel}

\affiliation[a]{SISSA - International School for Advanced Studies, Via Bonomea 265, 34136 Trieste, Italy}
\affiliation[b]{INFN – National Institute for Nuclear Physics, Via Valerio 2, I-34127 Trieste, Italy}
\affiliation[c]{IFPU, Institute for Fundamental Physics of the Universe, via Beirut 2, 34151 Trieste, Italy}
\affiliation[d]{INAF, Osservatorio Astronomico di Trieste, Via G. B. Tiepolo 11, I-34131 Trieste, Italy}
\affiliation[e]{ICSC - Italian Research Center on High Performance Computing, Big Data and Quantum Computing, via Magnanelli 2, 40033,
	Casalecchio di Reno, Italy}

\emailAdd{adambros@sissa.it}
\emailAdd{viel@sissa.it}

\abstract{Introducing primordial magnetic fields (PMFs) into the standard cosmological model, $\Lambda$CDM, has an impact on the matter power spectrum at the linear level, resulting in an increase of power on small scales. In this work, we use the enhanced spectrum as the initial condition for dark-matter-only N-body simulations and study its non-linear evolution. From the resulting non-linear matter power spectrum, we calculate the convergence power spectrum of weak lensing ($C_{\kappa}$) and show how the presence of PMFs causes an enhancement at high multipoles. We then construct an emulator for $C_{\kappa}$ that also includes intrinsic alignment and baryon feedback. We perform MCMC analyses using mock data generated for different PMF models and for three values of the maximum multipole, $\ell_{\rm max}=5000$, $10000$, and $20000$. The PMFs models are characterized by the two parameters that define the power-law power spectrum of the magnetic field $P_B(k)\propto B_{\rm 1Mpc}^2 k^{n_B}$, where $B_{\rm 1Mpc}$ is the amplitude at a scale of $1$ Mpc and $n_B$ is the power spectral index. For $\Lambda$CDM mocks, corresponding to $B_{\rm 1Mpc}=0$, we obtain approximate upper limits of $B_{\rm 1Mpc}<1.6$, $1.0$, and $0.8$ nG at the $95\%$ confidence level for the three values of $\ell_{\rm max}$, respectively. For non-zero input magnetic fields, the recovery of $B_{\rm 1Mpc}$ and $n_B$ depends strongly on the PMFs model and on the maximum multipole considered. Models with $n_B=-2.1$ are more easily constrained than models with $n_B=-2.7$, since they produce a stronger enhancement of the convergence power spectrum. We find no clear degeneracy between the PMFs parameters and the intrinsic alignment or baryon feedback parameters, while $B_{\rm 1Mpc}$ and $n_B$ can be degenerate. We characterize this degeneracy through linear fits to selected $2$D posterior distributions.}

\begin{document}
	\maketitle
	\flushbottom
	
\section{Introduction}\label{sec:intro}
Magnetic fields are observed on every scale probed so far, from the small scales of planets and stars to the large scales of galaxies, galaxy clusters and, tentatively, the voids of the cosmic web \cite{Kandus:2010ye,Durrer:2013pga,Subramanian:2015lua}. Although in the case of galaxies and galaxy clusters the origin of the magnetic fields can be attributed to astrophysical phenomena, later amplified by dynamo processes, their detection in cosmic voids \cite{doi:10.1126/science.1184192,Broderick_2012,Tjemsland:2023hmj,Vovk:2023qfk,MAGIC:2022piy,AlvesBatista:2021sln,VERITAS:2017gkr,Finke:2015ona,HESS:2014kkl,Vernstrom:2021uoi} is difficult to explain through late-time magnetogenesis alone and points instead
towards a primordial origin. Several mechanisms have been proposed to generate primordial magnetic fields (PMFs) in the early universe, including inflationary models  \cite{Sharma:2018kgs,Durrer:2010mq,Kobayashi:2014sga,PhysRevD.37.2743,Ratra:1991bn} and cosmological phase transitions \cite{Di:2020kbw,Yang:2021uid,Olea-Romacho:2023rhh,Sigl:1996dm,VACHASPATI1991258,Quashnock:1988vs}; a recent study of the turbulent decay of helical
fields generated at the electroweak phase transition has shown that such relics could be strong
enough to simultaneously explain the void observations and help alleviate the Hubble tension \cite{Hosking:2022jql}. Hints of PMFs have been found in early-universe observables, such as the cosmic microwave background (CMB) \cite{LiteBIRD:2024twk,Paoletti:2019pdi,Planck:2015zrl,Jedamzik:1999bm,Wagstaff:2015jaa,Paoletti:2022gsn,Trivedi:2018ejz}, the redshifted 21-cm line \cite{Bhaumik_2025}, and the intergalactic medium \cite{Natwariya:2020ksr,Sethi_2009,Sethi:2004pe}. \\
The introduction of PMFs in the cosmological model opens the possibility to improve our understanding of the universe, addressing some of the open problems of $\Lambda$CDM. One example is the Hubble tension \cite{Jedamzik:2020krr,Hosking:2022jql}, which PMFs could relieve through the modification of the size of the sound horizon at recombination. PMFs have also been discussed in connection with the $\sigma_8$-$S_8$ tension between early- and late-time measurements of the amplitude of matter clustering \cite{Abdalla:2022yfr}, a tension that is itself most directly probed by weak lensing and galaxy clustering surveys such as KiDS-1000 \cite{Asgari:2020wuj} and the Dark Energy Survey \cite{DES:2021wwk}. \\
Constraints on the strength of magnetic fields were found by studying the intergalactic medium \cite{MAGIC:2022piy}, filaments \cite{Carretti:2022fqk,Vernstrom:2021uoi}, and cosmic voids \cite{doi:10.1126/science.1184192,2024arXiv241214825N}, giving upper limits of the order of nG and lower limits of $10^{-8}$nG. Comparable limits, of a few nG on Mpc scales, come from Faraday rotation measurements of extragalactic radio sources \cite{OSullivan:2020bqx,Amaral:2021rgu}, while big bang nucleosynthesis provides an independent, much weaker bound on the field amplitude at early times \cite{Kawasaki:2012wb}. Another way to constrain PMFs parameters is to study observables affected by the increase in the linear matter power spectrum due to PMFs \cite{1978ApJ...224..337W,Kim:1994zh,Subramanian:1997gi}, such as the Lyman-$\alpha$ forest \cite{pandey2013probing,Chongchitnan:2013vpa}, dwarf galaxies \cite{sanati2024dwarf}, stellar mass density \cite{Zhang:2024yph} and reionization \cite{Tashiro:2005ua}. \\
In this work, we aim to forecast constraints on the strength and the power spectral index of PMFs by exploring their effect at the non-linear level. Cosmological magneto-hydrodynamic (MHD) simulations have traced the amplification and non-linear evolution of primordial and other seed magnetic fields through structure formation \cite{Marinacci:2015rga,Mtchedlidze:2021qzz}, and dedicated MHD simulations of the baryon perturbations directly sourced by PMFs have recently shown that the transition from the linear to the non-linear regime departs significantly from previous analytic expectations \cite{Ralegankar:2024dpb}. Here we follow a complementary, and computationally cheaper, approach: we evolve the PMF-induced matter power spectrum into the non-linear regime with dark-matter-only N-body simulations, seeded by the linearly evolved PMF and inflationary perturbations. A recent work \cite{10.1093/mnras/stae1848} showed how future cosmic shear surveys could be used to reconstruct the matter power spectrum on non-linear scales, so we compute a theoretical prediction for the convergence power spectrum of weak lensing in the presence of PMFs, which is equivalent to the shear power spectrum \cite{2017SchpJ..1232440B}. \\
We start from the evolution in the post-recombination era of baryon and dark matter perturbations due to the presence of a magnetic field. To be able to neglect any turbulent behavior, we focus on scales larger than the magnetic damping scale, an approximation that we refer to as the \textit{linear regime}. To compute the total matter perturbations, we add to these perturbations the contribution coming from inflation. Assuming a power-law shape for the magnetic field power spectrum, $P_B(k)\propto B_{\rm 1Mpc}^2 k^{n_B}$, we calculate the matter power spectrum for different values of the parameters that characterize PMFs, namely the strength $B_{\rm 1Mpc}$ and the power spectral index $n_B$, and use it to generate initial conditions at redshift $z=299$ for dark-matter-only N-body simulations. We run the simulations using P-GADGET3 to obtain the result of the non-linear evolution of matter perturbations. Extracting the non-linear matter power spectrum, necessary to study weak lensing \cite{Kilbinger:2014cea,Bartelmann:1999yn}, we perform a tomographic analysis of the convergence power spectrum, in which we include baryon feedback and intrinsic alignment. Baryon feedback is an important physical ingredient, since it is expected to suppress power at small scales due to galactic winds and/or Active Galactic Nuclei and increase power at very small scales due to star formation \cite{van_Daalen_2011, Semboloni_2011, van_Daalen_2019, chisari19, Schneider_2019}. This modulation of the matter power could in principle hide a PMF contribution or, if present at very small scales, increase the power due to star formation: one of the goals of this work is to exploit the mutual interplay between baryon effects and PMFs.
\\
The paper is organized as follows. In Section \ref{sec:theory} we show how we obtain the solution for the evolution of baryon and dark matter perturbations and how we calculate the linear matter power spectrum. In Section \ref{sec:WL} we describe how we construct the initial conditions for N-body simulations, how we obtain the non-linear matter power spectrum, and how we implement the calculation of the convergence power spectrum of weak lensing considering baryon feedback (BF) and intrinsic alignment (IA). We also explain how we construct an emulator of $C_{\kappa}(\ell;[B_{\rm 1Mpc}, n_B, A_{IA}, \log M_c, \eta_b])$ to use in the MCMC analysis. In Section \ref{sec:mcmc results} we present the results of the MCMCs we run on data mocks of the convergence power spectrum. We summarize our conclusions in Section \ref{sec:conclusions}.
	
\section{Linear theory}\label{sec:theory}

We consider only the post-recombination universe, where the baryon fluid can be treated as a perfect conductor, and so the magnetic field evolves following
\begin{align}\label{eq:B evol}
    \frac{\partial \vec{B}}{\partial t} = \frac{1}{a} \nabla \times (\vec{v}_b \times \vec{B}).
\end{align}
$\vec{B}$ is the comoving magnetic field, defined as $\vec{B}=a^2 \vec{B}_{\rm phys}$, $\vec{v}_b$ is the baryon bulk velocity, $a$ is the scale factor, set at unity today, and $t$ is physical time. \
On sub-horizon scales, the motion of baryons due to the Lorentz force is described by
\begin{align}\label{eq:euler}
    \frac{\partial \vec{v}_b}{\partial t} + H \vec{v}_{\rm b} + \frac{(\vec{v}_{\rm b} \cdot \nabla) \vec{v}_{\rm b}}{a} + \frac{c^2_{\rm b}}{a} \nabla \delta_{\rm b} = \frac{(\nabla \times \vec{B}) \times \vec{B}}{4 \pi a^5 \rho_{\rm b}} - \frac{\nabla \phi}{a},
\end{align}
where $H$ is the Hubble rate in physical time, $\rho_{\rm b}$ is the baryon energy density, $c_{\rm b}$ is the baryon sound speed, $\delta_{\rm b (DM)}$ is the baryon (dark matter) density  defined as $\delta_{\rm b (D)M } = \frac{\rho_{\rm b (DM)} - \overline{\rho}_{\rm b (DM)}}{\overline{\rho}_{\rm b (DM)}}$, and $\phi$ is the potential of the metric that satisfies the Poisson equation
\begin{align}\label{eq:poisson}
    \nabla^2 \phi = \frac{1}{2 M_{\rm Pl}^2} a^2 (\rho_b \delta_b + \rho_{\rm DM } \delta_{\rm DM}).
\end{align}
Here $M_{\rm Pl}=2.435 \times 10^{18}$ GeV is the reduced Planck mass. In the dark matter case, the sound speed and magnetic field are zero. \\
To describe the evolution of density perturbations, we introduce the continuity equation
\begin{align}\label{eq:continuity}
    \frac{\partial \delta}{\partial t}+\frac{\vec{\nabla}\cdot\vec{v}}{a}+\frac{\vec{\nabla}\cdot(\delta \vec{v})}{a}=0.
\end{align}
On large enough scales, we can neglect turbulence. To find this scale, called the damping scale $\lambda_{\rm D}$, we note that the baryon flow is driven by the Lorentz force, so
\begin{align}\label{eq:large scales}
    \sqrt{\langle v_{\rm b}^2\rangle} \sim \frac{1}{aH\lambda_{\rm D}}\frac{\langle\vec{B}^2\rangle a^{-4}}{4\pi\rho_{\rm b}}\equiv \frac{1}{aH\lambda_{\rm D}}v_A^2,
\end{align}
where $v_A$ is the Alfven velocity. Using $\rho_{\rm m0} \approx 1.15 \times 10^{-47} \text{GeV}^4$, $\rho_{\rm b0} \approx 1.64 \times 10^{-48} \text{GeV}^4$, and requiring $v_b/\lambda_D \sim aH$, we obtain
\begin{align}\label{eq:damping scale}
    \lambda_{\rm D}\sim \frac{v_{\rm A}}{aH}\approx 0.1{\rm Mpc} \left(\frac{\sqrt{\langle B^2\rangle}}{\rm nG}\right).
\end{align}
On scales larger than $\lambda_{\rm D}$, which we call the \textit{linear limit}, we can neglect the RHS of eq.~\eqref{eq:B evol} and the convective term $((\vec{v}_{\rm b} \cdot \nabla)\vec{v}_{\rm b})$ in eq.~\eqref{eq:euler}. In this regime, the magnetic field is constant in time.
\subsection{Evolution of baryon and dark matter perturbations}\label{subsec:delta evol}
\begin{figure}
    \centering
    \includegraphics[width=0.75\textwidth]{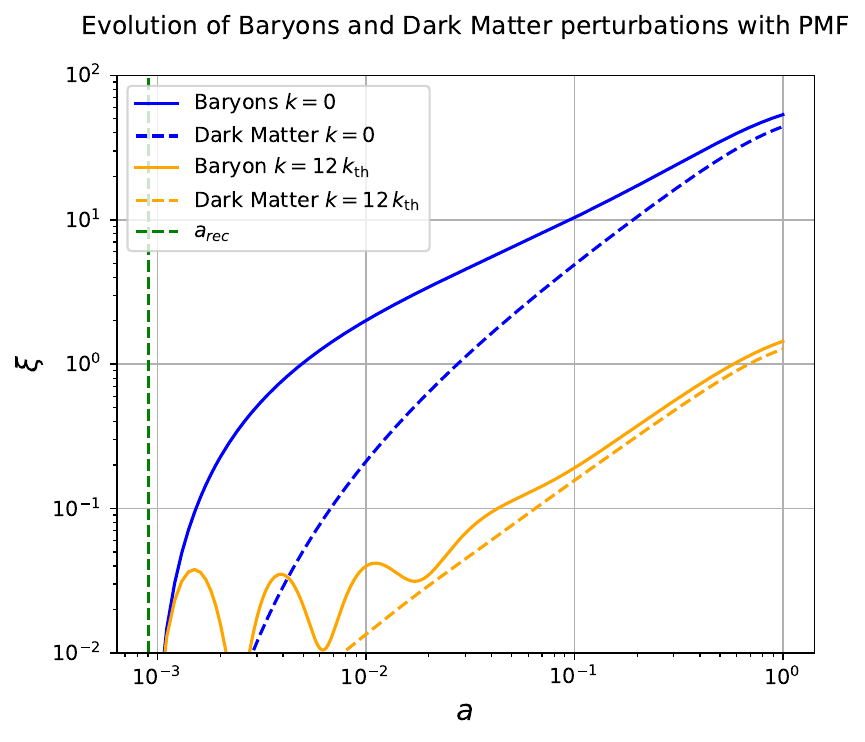}
    \caption{Evolution of baryon (solid line) and dark matter (dashed line) perturbations in the case of large scales (blue) and small scales (orange). For dark matter there are always two regimes where perturbations grow, first driven by the Lorentz force and then by gravity. In the small scale case, for baryons the first growing phase is followed by an oscillatory phase, during which the thermal pressure is of the same order of the Lorentz force; then the perturbations start to grow again thanks to gravity.}
    \label{fig:perturbation evol}
\end{figure}

Combining eq.~\eqref{eq:euler}, ~\eqref{eq:poisson} and ~\eqref{eq:continuity} in Fourier space, we can write the equations that describe the evolution of baryon and dark matter perturbations:
\begin{align}\label{eq:delta_b evol}
    a^2 \frac{\partial^2 \delta^{\rm lin}_b }{\partial a^2} + a \frac{3}{2}\frac{\partial \delta^{\rm lin}_b}{\partial a} + \frac{c^2_b}{(aH)^2} k^2 \delta^{\rm lin}_b - \frac{3}{2} \frac{\Omega_b}{\Omega (a)} \delta^{\rm lin}_b = - \left[ \frac{3 M^2_{Pl} S_B}{\rho_{m0}} \right] \frac{\Omega_m}{\Omega (a)} + \frac{3}{2}\frac{\Omega_{DM}}{\Omega (a)} \delta^{\rm lin}_{DM}
\end{align}
\begin{align}\label{eq:delta_DM evol}
    a^2 \frac{\partial^2 \delta^{\rm lin}_{DM} }{\partial a^2} + a \frac{3}{2}\frac{\partial \delta^{\rm lin}_{DM}}{\partial a} - \frac{3}{2} \frac{\Omega_{DM}}{\Omega (a)} \delta^{\rm lin}_{DM} = \frac{3}{2}\frac{\Omega_b}{\Omega (a)} \delta^{\rm lin}_b.
\end{align}
$\Omega(a) \equiv \Omega_m (1 + a_{eq} / a) + \Omega_{\Lambda} a^3$, with $\Omega_m= \Omega_b + \Omega_{DM} = 0.308$ and $\Omega_{\Lambda}=0.691$, where $a_{eq}$ is the value of the scale factor at the matter-radiation equality. The superscript "lin" indicates that these equations are valid in the linear limit described above. The entire dependence on the magnetic field is in the factor
\begin{align}\label{eq:S_B}
    S_B = \frac{\nabla \cdot \left[ \left( \nabla \times \overrightarrow{B}^{\rm lin} \right) \times \overrightarrow{B}^{\rm lin} \right]}{4 \pi \rho_{b0}},
\end{align}
which is constant in time. \\
In the linear regime of the system of ordinary differential equations that describe the perturbations evolution, the solution can be divided into the contribution due to inflation and the contribution due to PMFs. As a result, we can solve for $\left[ \delta^{\rm lin} \right]_{PMF}$ and then reconstruct the entire solution by adding the inflationary solution,
\begin{align}\label{eq:delta pmf+LCDM}
    \delta^{\rm lin} = \left[ \delta^{\rm lin} \right]_{PMF} + \left[ \delta^{\rm lin} \right]_{\Lambda CDM}.
\end{align}
To solve for $\delta_{\rm b}$ and $\delta_{\rm DM}$, we set their values and the values of their derivatives with respect to the scale factor to zero at recombination, so at $z=1100$. This choice does not affect the solution because the solution can be shown to be an attractor \cite{Ralegankar:2024dpb}. \\
One can isolate the time evolution of the density perturbation as
\begin{align}\label{eq:xi_b}
    \delta^{\rm lin}_b = - \xi_{\rm b} (k,a) \frac{3M_{Pl}^2}{\rho_{\rm m0}} S_B(k)
\end{align}
\begin{align}\label{eq:xi_DM}
    \delta^{\rm lin}_{DM} = - \xi_{\rm DM} (k,a) \frac{3M_{Pl}^2}{\rho_{\rm m0}} S_B(k)
\end{align}
where $\xi_{\rm b,DM}$ are independent of the strength of the magnetic field. \\
The behavior of the solution is different depending on the relation between the scale $k$ and the thermal scale
\begin{align}\label{eq:thermal scale}
    \lambda_{\rm th} = k_{\rm th}^{-1} = \sqrt{\frac{2 \Omega_m}{3 \Omega_b}} \frac{c_b}{aH}.
\end{align}
For scales $\lambda \gg \lambda_{\rm th}(a_{\rm rec})$, the baryon thermal pressure is negligible, so the evolution is first driven by the Lorentz force and then by gravity. For smaller scales, the baryon pressure becomes important, and when it is of the same order as the Lorentz force, the baryon perturbations start to oscillate until gravity prevails and baryon perturbations grow again. \\
We solve numerically the system of coupled differential equations eq.~\eqref{eq:delta_b evol} and ~\eqref{eq:delta_DM evol} for two scales, belonging to the two scale regimes described above, and plot them in Figure~\ref{fig:perturbation evol}.
\subsection{Linear matter power spectrum}\label{subsec:linear power spectrum}
\begin{figure}
    \centering
    \includegraphics[width=0.9\textwidth]{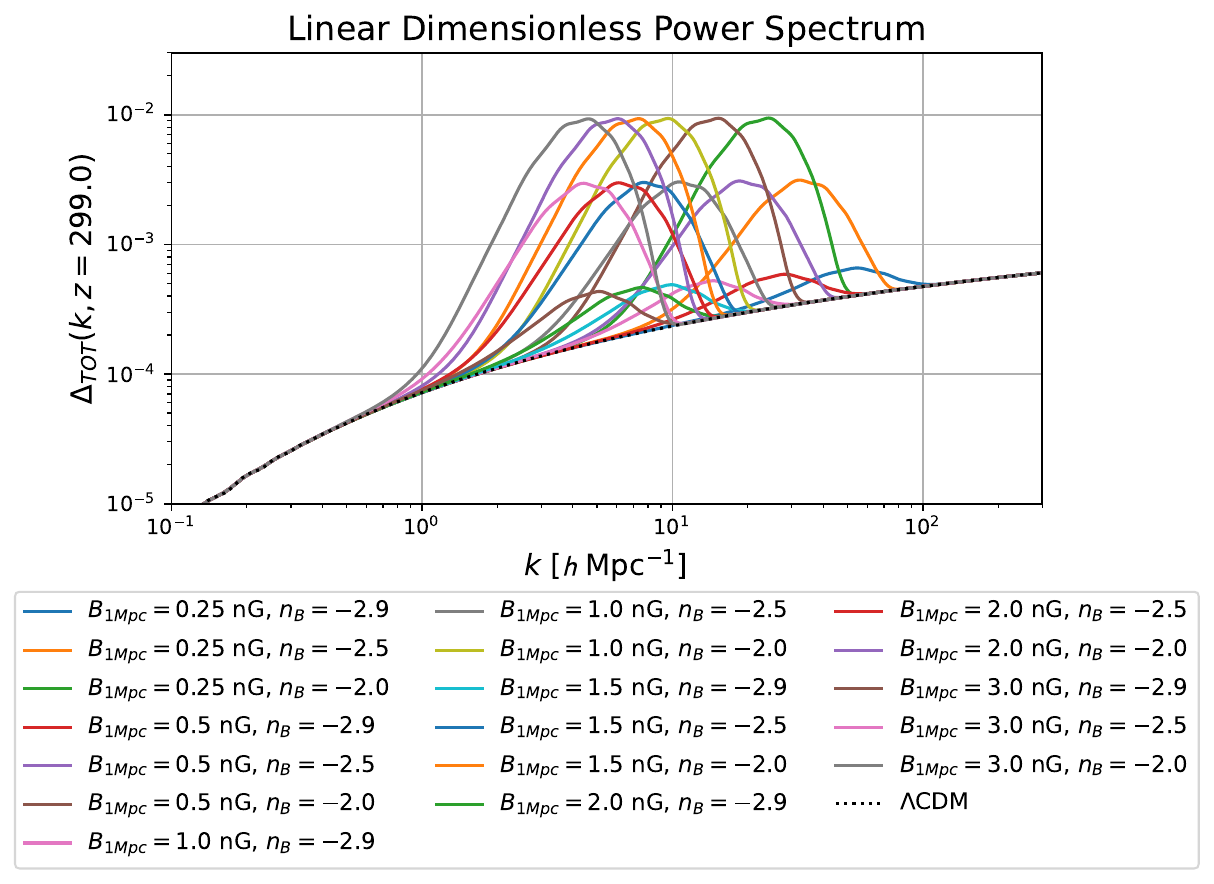}
    \caption{Dimensionless linear matter power spectrum at $z=299$ for different combinations of PMFs parameters and $\eta=0.1$ (colored solid line) and for $\Lambda$CDM (black dotted line), calculated with CAMB. The total power spectrum in the presence of PMFs is given by the sum of the power spectrum due to PMFs only and the power spectrum coming from the linear solution of inflationary perturbations.}
    \label{fig:dimensionless matter power spectrum}
\end{figure}

Through eq.~\eqref{eq:S_B} we can find an expression for the baryon and dark matter power spectrum, given an assumption on the magnetic field power spectrum. We treat non-helical magnetic fields, and we assume that their power spectrum is a power law
\begin{align}\label{eq:P_B(k)}
    P_B(k) = A k^{n_B} e^{-k^2 \lambda_D^2}.
\end{align}
The amplitude $A$ is set by the strength of the magnetic field at a scale of $\lambda_{\rm Mpc}=1$ Mpc through
\begin{align}\label{eq:B1Mpc}
    B^2_{1Mpc} \equiv \int \frac{d^3k}{(2\pi)^3} P^{\text{lin}}_B (k) e^{-k^2 \lambda^2_{Mpc}} = \frac{A \lambda^{-(3 + n_B)}_{Mpc}}{4 \pi^2} \Gamma \left( \left[ n_B + 3 \right] /2 \right),
\end{align}
where $\Gamma$ is the gamma function. $\lambda_{\rm D}$ is the cut-off scale defined in eq.~\eqref{eq:damping scale}, due to non-linear effects, which we included to be consistent with previous works \cite{Sethi:2004pe,Kim:1994zh,Adi:2023doe,Finelli:2008xh,Paoletti:2008ck,Kunze:2021qxt}. Taking the ensemble average $\langle \delta(k) \delta^*(k') \rangle$, and using the ensemble average of the magnetic field
\begin{align}\label{eq:B ensemble average}
    \langle B_i(k) B_j^*(k') \rangle = (2\pi)^3 \delta^3 (k-k') \left( \delta_{ij} - \frac{k_i k_j}{k^2} \right) \frac{P_B (k)}{2},
\end{align}
we can write the expression for the power spectrum of baryon and dark matter in the linear limit:
\begin{align}\label{eq:linear power spectrum}
    P^{\rm lin} (k) = & \xi^2 (k,a) \frac{9k^4 M^4_{\rm Pl}}{8 (4 \pi \rho_{\rm b0} \rho_{\rm m0})^2} \int \frac{d^3 q}{(2 \pi)^3} \frac{P_B(q)P_B(k-q)}{(k-q)^2} \left[ k^2 + 2q^2 + 4\frac{(q \cdot k)^4}{k^4 q^2} \right. \nonumber \\
    & \left. -4\frac{(q \cdot k)^2}{k^2} -4\frac{(q \cdot k)^3}{k^2 q^2} + \frac{(q \cdot k)^2}{q^2} \right].
\end{align}
We consider only values of the power spectral index $n_B$ in the range $[-3,-1.5]$, so that the integral in eq.~\eqref{eq:linear power spectrum} is independent of the cut-off of both small and large scales, resulting in $P_B(k)=Ak^{n_B}$. In this regime, the dimensionless power spectrum is
\begin{align}\label{eq:adimensional power spectrum}
    \Delta^{\text{lin}} (k) = \frac{k^3 P^{\text{lin}} (k)}{2 \pi^2} \approx 0.918 \times 10^{-4} \xi^2 (k,a) \left ( \frac{k}{Mpc^{-1}} \right)^{2n_B + 10} \left( \frac{B_{\rm 1Mpc}}{nG} \right)^4 G_{n_B},
\end{align}
where we use $\rho_{\rm m0} \approx 1.15 \times 10^{-47} \text{GeV}^4$ and $\rho_{\rm b0} \approx 1.64 \times 10^{-48} \text{GeV}^4$, and $G_{n_B}$ is given by
\begin{align}\label{eq:G_nB}
    G_{n_B} = \int_0^{\infty} dx \int_{-1}^{1} \frac{dy}{2} x^{n_B + 2} \left( 1 + x^2 -2xy \right)^{n_B/2 -1} \times \frac{\left[ 1 + 2x^2 + 4y^4x^2 - 4y^2x^2 - 4y^3x + y^2 \right]}{\Gamma^2 \left( \left[n_B + 3 \right]/2 \right)}.
\end{align}
We want to extrapolate the linear results to scales where the non-linear effects cannot be neglected, and following earlier works \cite{Sethi:2004pe,Kim:1994zh,Shaw_2012,Adi:2023doe,Finelli:2008xh,Paoletti:2008ck,Kunze:2021qxt}, we introduce the damping term
\begin{align}\label{eq:damping factor}
E_{n_B}(z) = & \frac{1}{G_{n_B}} \int_0^{\infty} dx \int_{-1}^{1} \frac{dy}{2} \left[ x^{n_B + 2} \left( 1 + x^2 -2xy \right)^{n_B/2 -1} e^{-\left( 1 + 2x^2 - 2xy \right)z^2} \right. \nonumber \\
& \left. \times \frac{\left[ 1 + 2x^2 + 4y^4x^2 - 4y^2x^2 - 4y^3x + y^2 \right]}{\Gamma^2 \left( \left[n_B + 3 \right] /2 \right) }\right] ,
\end{align}
that multiplies the linear power spectrum, giving
\begin{align} \label{eq:damped power spectrum}
    \Delta(k) = \Delta^{\text{lin}} (k) E_{n_B} (k \lambda_D).
\end{align}
From eq.~\eqref{eq:damping scale} we can see that there is a direct proportionality between the damping scale $\lambda_{\rm D}$ and the magnetic field $B$, which can be parametrized through
\begin{align}\label{eq:lambdaD B proportionality}
    \lambda_{\rm D} = 0.1 \kappa_D(n_B) \text{Mpc} \left( \frac{B}{\text{nG}} \right).
\end{align}
The parameter $\kappa_D$ enters into the dimensionless baryon power spectrum when it is calculated at the damping scale as
\begin{align}\label{eq:dimensionless power spectrum at lambda_D}
    \Delta_{\rm b}^{\rm PMF} (k=\lambda_{\rm D}^{-1}) = \xi_{\rm b}^2 (a) \kappa_D^{-4} G_{n_B} e^{-2}.
\end{align}
We can encapsulate $\kappa_D$ into another parameter, $\eta$, corresponding to the value of $\Delta_{\rm b}^{\rm PMF} (k=\lambda_{\rm D}^{-1},a=0.01)$. Considering that when $\Delta_{\rm b}^{\rm PMF}$ is of order unity, so when baryon fluctuations become non-linear, $\xi_b \approx 2.5$, we can fix $\xi_b (a=0.01)=2.5$ and write the relation between $\kappa_D$ and $\eta$, obtaining
\begin{align}\label{eq:kappa eta proportionality}
    \kappa_D(n_B) = \left( \frac{G_{n_B}}{1.14 \eta} \right)^{1/4}.
\end{align}
To avoid numerical artifacts that cause the damping of the power spectrum, we choose $\eta=0.1$. \\
The matter perturbations are defined as the sum of baryon and dark matter perturbations, weighted by their density with respect to the total matter density, $\xi_m = \xi_b \Omega_b / \Omega_m + \xi_{DM} \Omega_{DM} / \Omega_m$, with $\Omega_m = 0.308$ and $\Omega_b=0.0482$. We compute the dimensionless matter power spectrum for different combinations of the PMFs parameters $B_{1\text{Mpc}}$ and $n_B$ at redshift $z=299$ and report the results in Figure \ref{fig:dimensionless matter power spectrum}. The power spectrum due to the inflationary solution is calculated using the Einstein-Boltzmann code CAMB \cite{Lewis_2000}, so that it can be added to the power spectrum due to PMFs to obtain the total solution, as defined in eq.~\eqref{eq:delta pmf+LCDM}. \\
The presence of PMFs results in an increase of the matter power spectrum at small scales, in the range $1-100  \ h/\text{Mpc}$. The position and height of the bump depend on the parameters chosen.

\section{Non-linear evolution}\label{sec:WL}
In the following section, we describe how we use the linear matter power spectrum derived in Section \ref{subsec:linear power spectrum} as initial conditions (ICs) in dark-matter-only N-body simulations to extract the non-linear power spectrum.

\subsection{N-body simulations}\label{subsec:IC}
\begin{figure}
    \centering
    \includegraphics[width=1\linewidth]{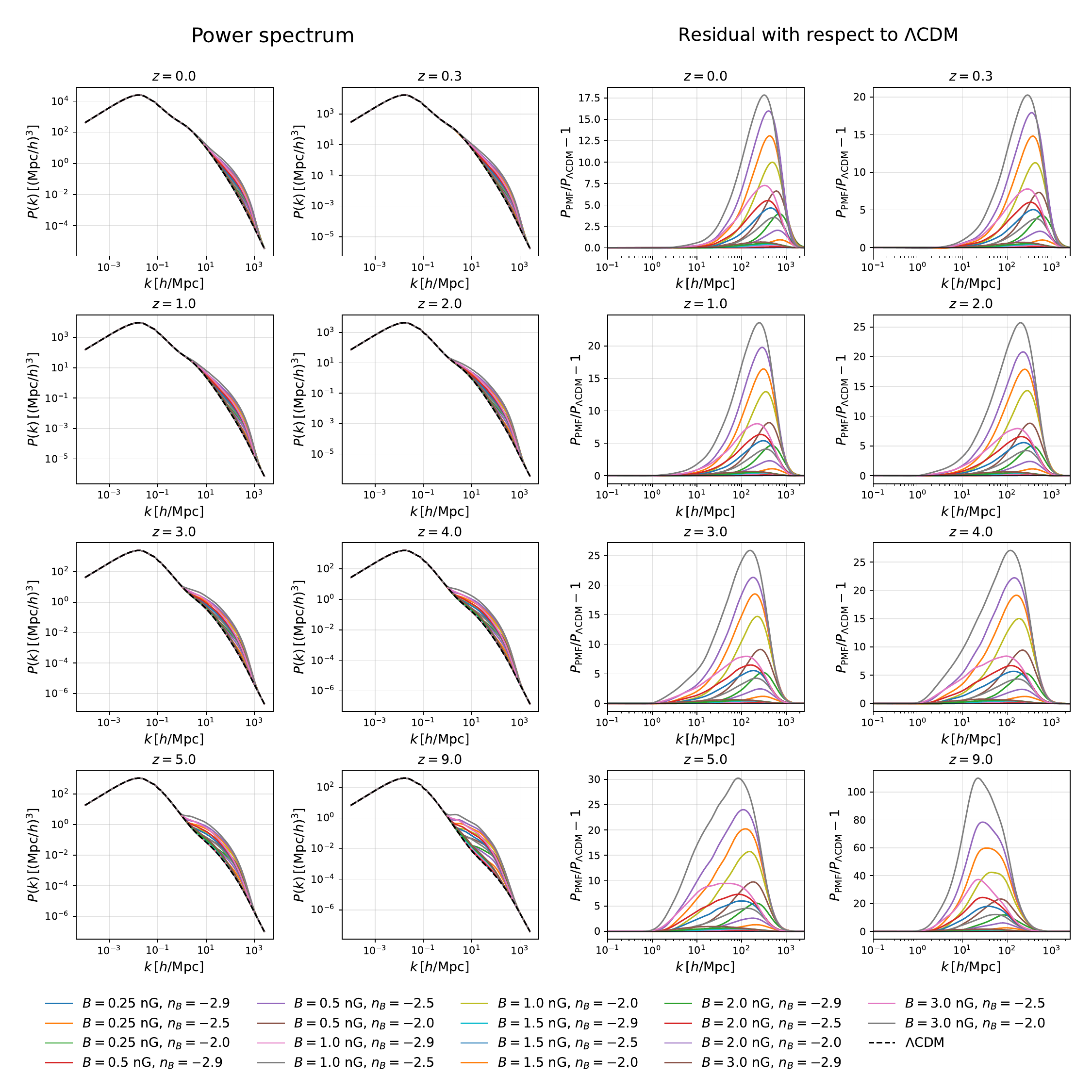}
    \caption{\textbf{Left}: Non-linear matter power spectrum taken from the snapshots of the N-body simulations for $\Lambda$CDM and the different PMFs parameter combinations. \textbf{Right}: Residual of the non-linear matter power spectrum in the presence of PMFs with respect to the $\Lambda$CDM case. At higher redshifts, the peaks are higher, and they are positioned at larger scales.}
    \label{fig:non-linear Pk and residual}
\end{figure}

For all the linear matter power spectra at $z=299$ in Figure \ref{fig:dimensionless matter power spectrum}, we calculate the ICs using N-GenIC \cite{Springel_2005, Angulo_2012}. In Appendix \ref{appendix:IC}, we show the power spectra of the ICs.
We run the simulations using the P-GADGET3 code, a non-public extension of the public GADGET2 code \cite{Springel:2005mi}. The box size is $40 \text{ Mpc}/h$, with $1024^3$ particles, resulting in a dark matter (DM) particle mass of $4.3 \cdot 10^{6}$ $M_{\odot}/h$. The gravitational softening is chosen to be 1/40 of the mean linear interparticle separation. The cosmological parameters of the ICs are in agreement with \textit{Planck} \cite{Planck:2015fie}: $\Omega_m=0.308$, $n_{\rm s}=0.968$, $\Omega_b=0.0482$, $h=0.678$, $\sigma_8=0.829$.
We run a total of $19$ simulations, for the following models: $\Lambda$CDM, and combinations of $B_{\rm 1Mpc}$ and $n_B$ with $B_{\rm 1Mpc}\in\{0.25, 0.5, 1.0, 1.5, 2.0, 3.0\}$nG and $n_B\in\{-2.9, -2.5, -2.0\}$.
For the PMFs simulations we also tested other box sizes, redshifts of the initial conditions and resolutions to check whether the non-linear power spectrum was numerically converged at the scales of interest for this work. The CPU time needed for each simulation is about 25,000 CPU hours. \\
We take snapshots at different redshifts: $9.0$, $5.0$, $4.0$, $3.0$, $2.0$, $1.0$, $0.3$, $0.0$, and analyze them with the software Pylians\footnote{\url{https://pylians3.readthedocs.io/en/master/}} \cite{Pylians} to extract the non-linear matter power spectrum. To mitigate the effects of cosmic variance, which affects the $\Lambda$CDM and PMFs simulations in the same way, and to expand the scale range to larger scales, not affected by the PMFs bump, we take the ratio of the PMF and $\Lambda$CDM power spectra and multiply this ratio by the non-linear matter power spectrum generated with CAMB. In Figure \ref{fig:non-linear Pk and residual} we report on the left panel the non-linear matter power spectrum and on the right the residual of the power spectrum with PMFs with respect to the $\Lambda$CDM power spectrum. Comparing the non-linear power spectra in Figure \ref{fig:non-linear Pk and residual} to the linear power spectra in Figure \ref{fig:IC}, we can note that the bumps are less pronounced, probably because the non-linear evolution mixes higher and lower scales, redistributing the initial enhancement across a wider range of scales. The position and height of the peaks differ based on the parameters considered. Going to higher redshifts, the peaks increase in amplitude and move to larger scales. \\
We then calculate the value of $\sigma_8$ for every combination of PMF parameters for which we run the N-body simulations, using the linear power spectra at $z=0$, through the same procedure used to calculate the ICs. The value of $\sigma_8$ increases for larger $B_{1\rm Mpc}$ and for $n_{\rm B}$ less negative. Our results are all compatible at $1\sigma$ with \textit{Planck2015} \cite{Planck:2015fie}, which found $\sigma_8=0.829 \pm 0.014$, except for [$B_{\rm 1Mpc}=3, n_B=-2.0$] for which we found $\sigma_8 = 0.851$. \\
\subsection{Modelling baryon feedback}\label{subsec:bf}
\begin{figure}
    \centering
    \includegraphics[width=0.75\linewidth]{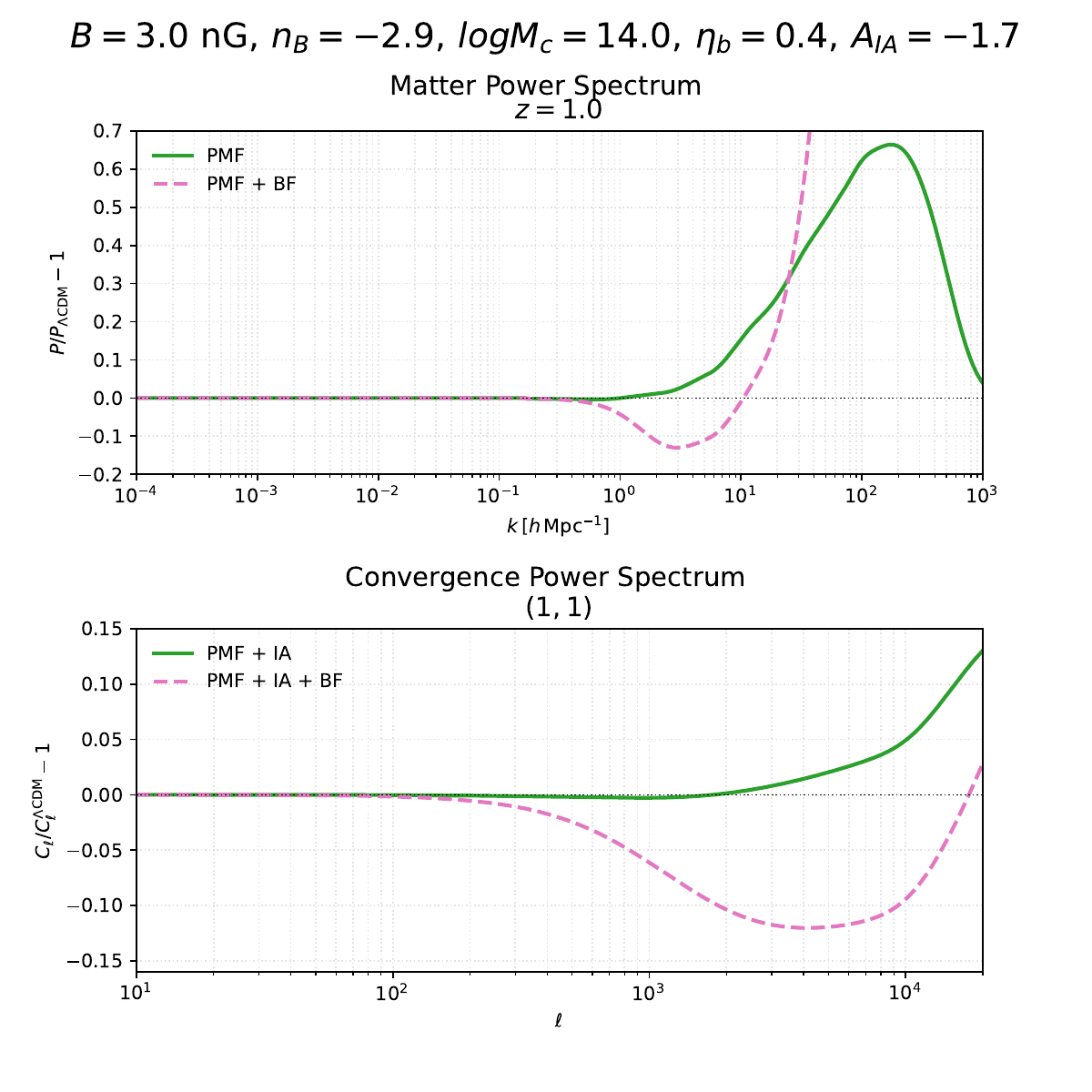}
    \caption{Comparison of the matter power spectrum at $z=1$ (upper subplot) and the convergence power spectrum in the $(1,1)$ bin (lower subplot) in the presence of PMFs with (pink dashed line) and without (green solid line) baryon feedback. Overall the matter power spectrum with baryon feedback at non-linear scales sees a first decrease in power and then an increase at smaller scales. In the convergence power spectrum we see a more pronounced dip when baryon feedback is applied and then an increase of power at smaller scales.}
    \label{fig:power spectra with baryon feedback}
\end{figure}

We introduce baryon feedback (BF) as a multiplicative factor of the non-linear matter power spectrum, and we use the result to calculate the convergence power spectrum of weak lensing.
To implement baryon feedback in our analysis, we use the \textit{baryon correction model} from \cite{Schneider:2015wta}, which was obtained from hydrodynamical simulations that incorporate AGN feedback but not other feedback mechanisms. The correction is applied at the level of the matter power spectrum ($P_{\rm m}$) with a scale and redshift-dependent multiplicative factor $F_{\rm bf}(k,z)$, including three free parameters: $M_c$, $\eta_b$, and $z_c$. $F_{\rm bf}$ is defined as
\begin{align}\label{eq:bf F}
    F_{\rm bf} (k, z | M_c, \eta_b, z_c) \equiv \frac{P_{\rm feed}(k)}{P_{\rm m}(k)} = \left\{ \frac{B(z)}{1 + (k/k_g)^3} + [1 - B(z)] \right\} S(k),
\end{align}
where
\begin{align}\label{eq:bf B}
    B(z) = \frac{0.105 \log \left ( \frac{M_c}{M_{\odot}/h} \right) - 1.27}{1 + (z/z_c)^{2.5}} \text{ if } M_c \geq 10^{12} M_{\odot} / h \text{ , zero otherwise},
\end{align}
\begin{align}\label{eq:bf kg}
    k_g (z) = 0.7 [1 - B(z)]^4 \eta_b ^{-1.6} h \text{Mpc} ^{-1},
\end{align}
and
\begin{align}\label{eq:bf S}
    S(k) = 1 + \left( \frac{k}{55 h \text{Mpc}^{-1}} \right)^2.
\end{align}
$M_c$ is the limit mass at which most gas is ejected, expressed in $M_{\odot}/h$; $\eta_b$ is the maximum scale of relevance for the suppression; $z_c$ describes the time dependence of the suppression, which is stronger at lower redshift. \\
In the top panel of Figure \ref{fig:power spectra with baryon feedback} we plot the residual of the non-linear matter power spectrum with PMFs with and without baryon feedback at $z=1$ with respect to the $\Lambda$CDM spectrum without BF. The addition of baryon feedback results in a dip whose position and depth depend on the values of $M_c$, $\eta_b$ and $z_c$; in particular an increase in $M_c$ or in $\eta_b$ moves the dip to smaller scales.
\subsection{Convergence power spectrum of weak lensing}\label{subsec:Ck}
We compute a theoretical prediction for the convergence power spectrum of weak lensing $C_{\kappa}(\ell)$ (\cite{Bartelmann:1999yn} for an extensive review), from now on denoted by $C(\ell)$ for brevity. We assume the flat-sky Limber approximation \cite{Kilbinger:2017lvu}. To improve the constraining power \cite{Stafford:2021uvk}, we perform a tomographic analysis using three redshift bins $[0.0, 0.7], [0.7, 1.11], [1.11, 5]$, which will be later referred to respectively with indices $0$, $1$ and $2$. Because the effects of PMFs on the matter power spectrum can be seen only on small scales, we choose to consider a large range of multipoles, $\ell  \in [10, 2 \cdot 10^4]$, even though it is larger than what surveys like \textit{Euclid} could optimistically probe \cite{Euclid:2019clj}. Following \cite{Parimbelli_2019}, we define $C^{(ij)}$, where the indices $ij$ denote the redshift bin combination considered:
\begin{align}\label{eq:C_k^ij}
    C^{(ij)} (\ell) = \int_0^{\infty} dz \frac{c}{H(z)} \frac{W^{(i)}(z) W^{(j)}(z)}{\chi^2(z)} P_m\left( k=\frac{\ell}{\chi(z)},z \right).
\end{align}
$W^{(i)}$ is the window function of the $i$-th bin defined as
\begin{align}\label{eq:Window}
    W^{(i)}(z)= \frac{3}{2} \Omega_m \left( \frac{H_0}{c} \right)^2 (1 + z) \chi(z) \int_{\text{max}(z,z_i)}^{z_{i+1}} dx \ n_s(x) \frac{\chi(x) - \chi(z)}{\chi(x)},
\end{align}
where $H_0$ is the Hubble function at present time, $c$ is the speed of light, $\chi$ is the comoving distance and $n_s$ is the distribution of source galaxies normalized to $1$. To calculate the window function, we need to assume a shape for $n_s$, for which we take
\begin{align}\label{eq:ns}
    n_s (z) = \frac{\beta}{z_0 \Gamma \left( \frac{\alpha + 1}{\beta} \right)} \left( \frac{z}{z_0} \right)^{\alpha} \exp \left[ -\left( \frac{z}{z_0} \right)^{\beta} \right]
\end{align}
with $\alpha=2$, $\beta=1.5$ and $z_0=0.9$ \cite{Euclid:2019clj}. The power spectra used are those in Figure \ref{fig:non-linear Pk and residual}, multiplied by the correction factor due to baryon feedback defined in eq.~\eqref{eq:bf F}. \\
When studying weak lensing, it is important to take into consideration intrinsic alignment (IA). The shapes of nearby sources are correlated because they form and evolve in the same gravitational environment, so in addition to the standard term ($C^{ij}_{\rm GG}$) of eq.~\eqref{eq:C_k^ij}, we need to add the correlation of shapes of nearby galaxies ($C^{ij}_{\rm II}$), and the correlation of galaxies aligned with those lensed by the same structure ($C^{ij}_{\rm GI}$). These are respectively defined by
\begin{align}\label{eq:IA C_II}
    C^{ij}_{\text{II}} (\ell) = \int_0^{\infty} dz \frac{c}{H(z)} \frac{N^2_s(z) }{\chi^2(z)} \times F^2_{IA} \left( k=\frac{\ell}{\chi(z)},z \right) P_m \left( k=\frac{\ell}{\chi(z)},z \right),
\end{align}
\begin{align}\label{eq:IA C_GI}
    C^{ij}_{\text{GI}} (\ell) = \int_0^{\infty} dz \frac{c}{H(z)} \frac{N_s(z) \left(W^{(j)} + W^{(i)} \right)}{\chi^2(z)} F_{IA} \left( k=\frac{\ell}{\chi(z)},z \right) P_m \left( k=\frac{\ell}{\chi(z)},z \right),
\end{align}
with
\begin{align}\label{eq:IA Ns}
    N_s(z) = n_s(z) \frac{H(z)}{c},
\end{align}
\begin{align}\label{eq:IA F_IA}
    F_{IA} (k,z) = -A_{IA} C_1 \rho_c \frac{\Omega_m}{D(z)}.
\end{align}
$\rho_c$ is the critical density today $\left( \rho_c = \frac{3H_0^2}{8 \pi G} \right)$, $D(z)$ is the growth factor, and $A_{IA}$ is the parameter that defines the strength of the effect of intrinsic alignment. $C_1=5 \times 10^{-14} h^{-2} M_{\odot}^{-1} \text{Mpc}^3$ is a normalization constant chosen so that $A_{IA}$ takes values around unity. The growth factor is calculated using the $\Lambda$CDM model, because in the approximation considered the magnetic field does not affect the background expansion. The complete signal is given by
\begin{align}\label{eq:C_tot}
    C^{ij}_{\text{tot}} (\ell) = C^{ij}_{\text{GG}} (\ell) + C^{ij}_{\text{II}} (\ell) + C^{ij}_{\text{GI}} (\ell).
\end{align}
To run the MCMC, we need the likelihood; in our case we use the \textit{log-likelihood}, defined as
\begin{align}\label{eq:log_likelihood}
    \log \mathcal{L} = - \frac{1}{2} \sum_{i,j} \sum_{\ell, \ell'} \left( C^{(ij)}_{\text{obs}} - C^{(ij)}_{\text{th} } \right)_{\ell} \left[ \mathcal{C}^{(ij)} \right]^{-1}_{\ell \ell'} \left( C^{(ij)}_{\text{obs}} - C^{(ij)}_{\text{th} } \right)_{\ell'},
\end{align}
where we sum over all possible redshift bin combinations $(ij)$ defined at the beginning of this section, and all multipoles $\ell$. $C_{\rm obs}$ is a data mock of the convergence power spectrum and $C_{\rm th}$ is the value of $C_{\kappa}$ calculated during the MCMC using eq.~\eqref{eq:C_tot}. To define the covariance $\mathcal{C}$, we need to take into consideration as sources of error: cosmic variance (CV), shape noise (SN) and the theoretical error on the matter power spectrum propagated in the covariance spectrum (sys), to obtain
\begin{align}\label{eq:Covariance_tot}
    \mathcal{C}^{(ij)}_{\ell \ell'} = \mathcal{C}^{(ij),\text{CV-SN}}_{\ell \ell'}+ \mathcal{C}^{(ij),\text{sys}}_{\ell \ell'}.
\end{align}
The sum of the contributions of cosmic variance and shape noise is given by
\begin{align}\label{eq:Cov_CV,SN}
    \mathcal{C}^{(ij),\text{CV-SN}}_{\ell \ell'}  = \frac{2}{2\ell + 1} f_{\text{sky}}^{-1} \left[ C^{(ij)}_{\ell} + N^{(ij)}_{\ell} \right]^2 \delta_{\ell \ell'},
\end{align}
where $C$ is the convergence power spectrum and $N$ is defined as
\begin{align}\label{eq:Cov N}
    N^{(ij)} (\ell) = \delta_{ij} \langle \gamma^2_{\rm rms} \rangle n^{-1}_i.
\end{align}
$f_{\rm sky}$ is the sky coverage of the hypothetical survey, $n_i$ is the number of galaxies per steradian in the $i$-th bin, $n_i = 3600 d (180/\pi)^2/N$, where $N$ is the number of bins, $d$ is the galaxy density and $\langle \gamma^2_{\rm rms} \rangle^{1/2}$ is the root mean square intrinsic shear. \\
The systematic error is defined as
\begin{align}\label{eq:Cov_sys}
    \mathcal{C}^{(ij),\text{sys}}_{\ell \ell'} = E^{(ij)}_{\ell} E^{(ij)}_{\ell'} \exp \left[ -\frac{\log^2 (\ell/\ell')}{2 \sigma^2_{\rm E}} \right]
\end{align}
where $E$ is the propagation of the error of the matter power spectrum through eq.~\eqref{eq:C_k^ij}, which results in
\begin{align}\label{eq:Cov E}
    E^{(ij)} (\ell) \equiv \frac{\Delta C^{(ij)}(\ell)}{C^{(ij)}(\ell)} = \int_0^{\infty} dz \frac{c}{H(z)} \frac{W^{(i)}(z)W^{(j)}(z)}{\chi^2(z)} \alpha \left( k=\frac{\ell}{\chi(z)},z \right) P_m \left( k=\frac{\ell}{\chi(z)},z \right).
\end{align}
$\sigma_{\rm E}$ is the correlation length of the systematic error, and $\alpha$ is the error associated with N-body methods to calculate the non-linear matter power spectrum. Following \cite{Bird_2012}, $\alpha$ takes the form
\begin{align}\label{eq:Cov alpha}
    \alpha \equiv \frac{\Delta P_m (k,z)}{P_m (k,z)} = \frac{\ln \left[ 1 + k/k_{\sigma} (z) \right]  }{1 + \ln \left[ 1 + k/k_{\sigma} (z) \right]} \alpha_{\text{max}},
\end{align}
which saturates at $\alpha_{\rm max}$ at large $k$. $k_{\sigma}$ is the scale at which mass fluctuations reach unity, so that $k_{\sigma} = \left. \frac{1}{R} \right|_{\sigma=1}$. Assuming again that in the approximation used here, PMFs do not modify the growth evolution, $k_{\sigma}$ is calculated using the $\Lambda$CDM model. \\
In Table \ref{tab:Cov params} we summarize the values chosen for the parameters that appear in the covariance, in accordance with \cite{Euclid:2019clj}. The effect of PMFs on the convergence power spectrum is rather small and can be seen only at very large multipoles, so the values of the covariance parameters are chosen accordingly, in order to be able to distinguish between different PMFs models.
\begin{table}
\begin{center}
\begin{tabular}{|c|c|}
\hline
$f_{\text{sky}}$ & $0.375$ \\
\hline
$N$ & $3$ \\
\hline
$d$ & $30$ $\text{arcmin}^{-2}$ \\
\hline
$\left< \gamma^2_{\text{rms}} \right>^{1/2}$ & $0.30$ \\
\hline
$\sigma_{\rm E}$ & $\log 3$\\
\hline
$\alpha_{\text{max}}$ & $0.025$ \\
\hline
\end{tabular}
\end{center}
\caption{Parameters used in the covariance of the convergence power spectrum.}
\label{tab:Cov params}
\end{table} \\
With the non-linear matter power spectrum extracted from the N-body simulations, we apply the baryon feedback and calculate the convergence power spectrum, also containing intrinsic alignment. We repeat this for all values of $B_{\rm 1Mpc}$ and $n_B$, and for different values of $A_{IA}$, $\log M_c$ and $\eta_b$, keeping $z_c$ fixed. Using multiquadric radial basis function interpolation, we construct an emulator for $C_{\kappa}(\ell;[B_{\rm 1Mpc}, n_B, A_{IA}, \log M_c, \eta_b])$. In the bottom panel of Figure \ref{fig:power spectra with baryon feedback} we show an example of the convergence power spectrum calculated using our emulator: we plot the residual of the spectrum with PMFs, with and without BF, with respect to the $\Lambda$CDM one. When BF is present, there is a more prominent decrease in power at small scales, and then an increase at smaller scales.
\section{Results}\label{sec:mcmc results}
\begin{figure}
    \centering
    \includegraphics[width=1\textwidth]{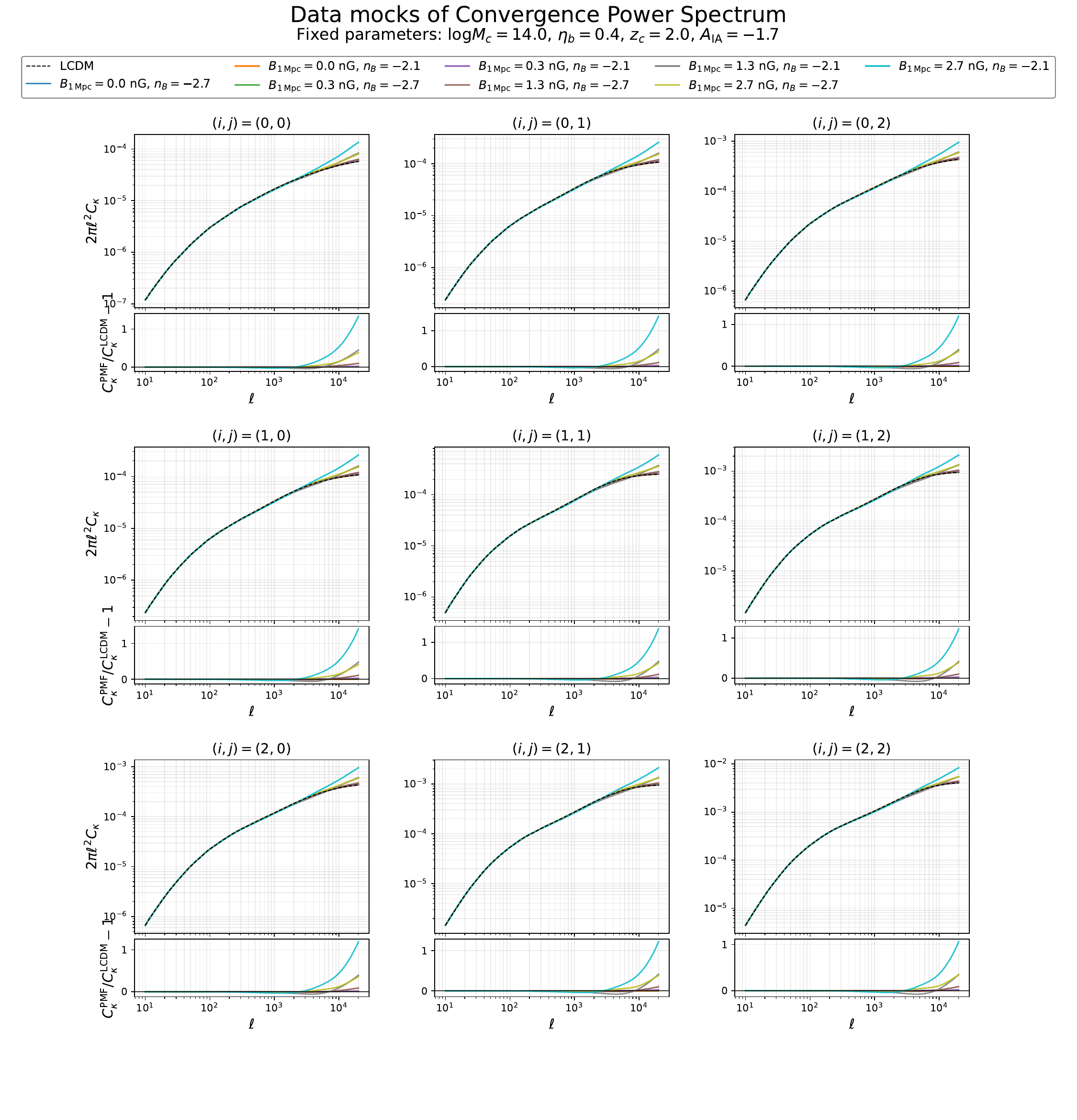}
    \caption{Data mocks of the convergence power spectrum created keeping all parameters fixed, varying only $B_{\rm 1Mpc}$ and $n_B$. Each subplot corresponds to a redshift bin combination, with on top the convergence power spectra and on the bottom their residual with respect to the $\Lambda$CDM model.}
    \label{fig:Ck data mocks}
\end{figure}

\begin{figure}[ht]
\centering
\begin{minipage}{0.49\textwidth}
\centering
\includegraphics[width=\linewidth]{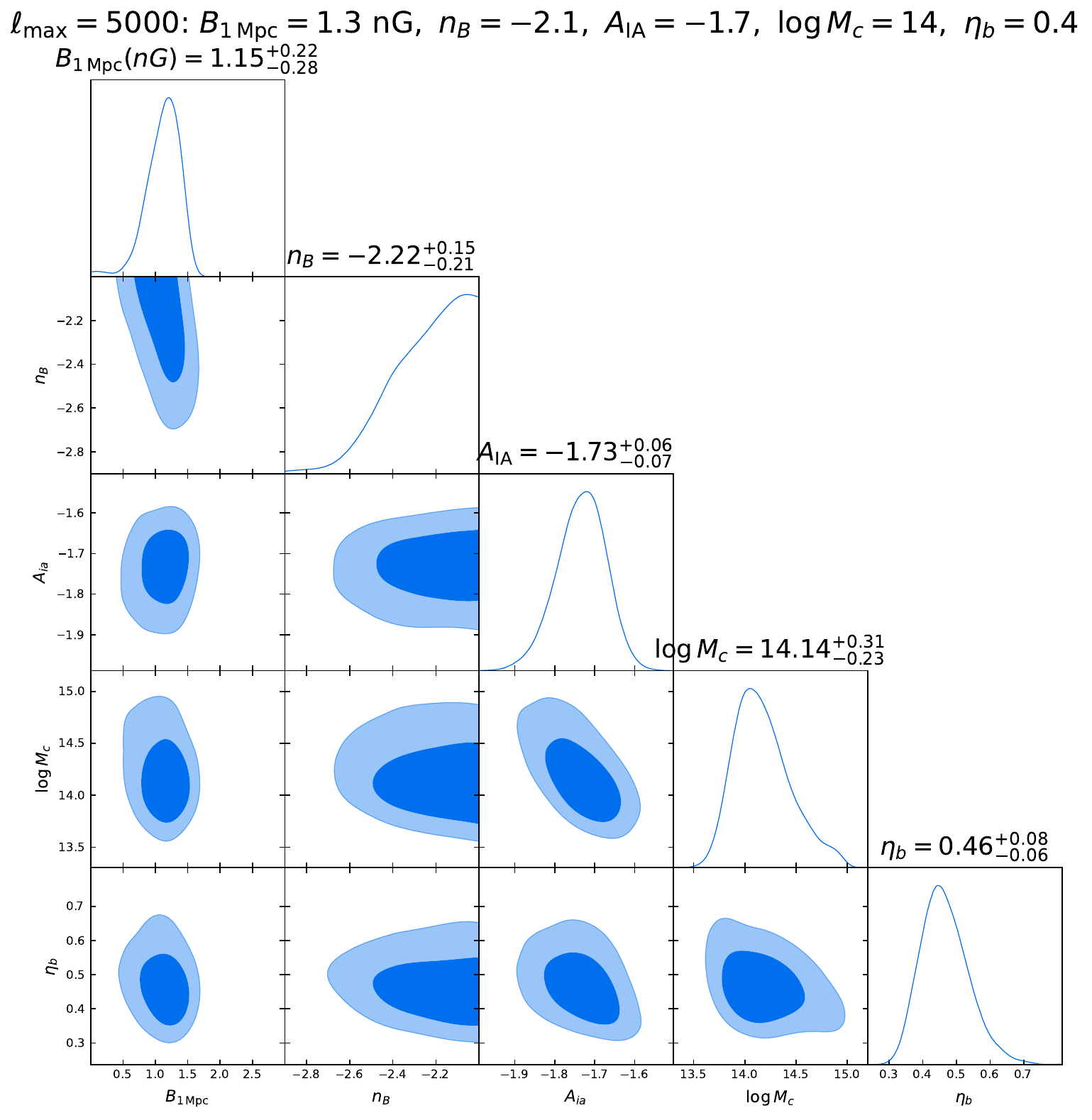}
\end{minipage}
\hfill
\begin{minipage}{0.49\textwidth}
\centering
\includegraphics[width=\linewidth]{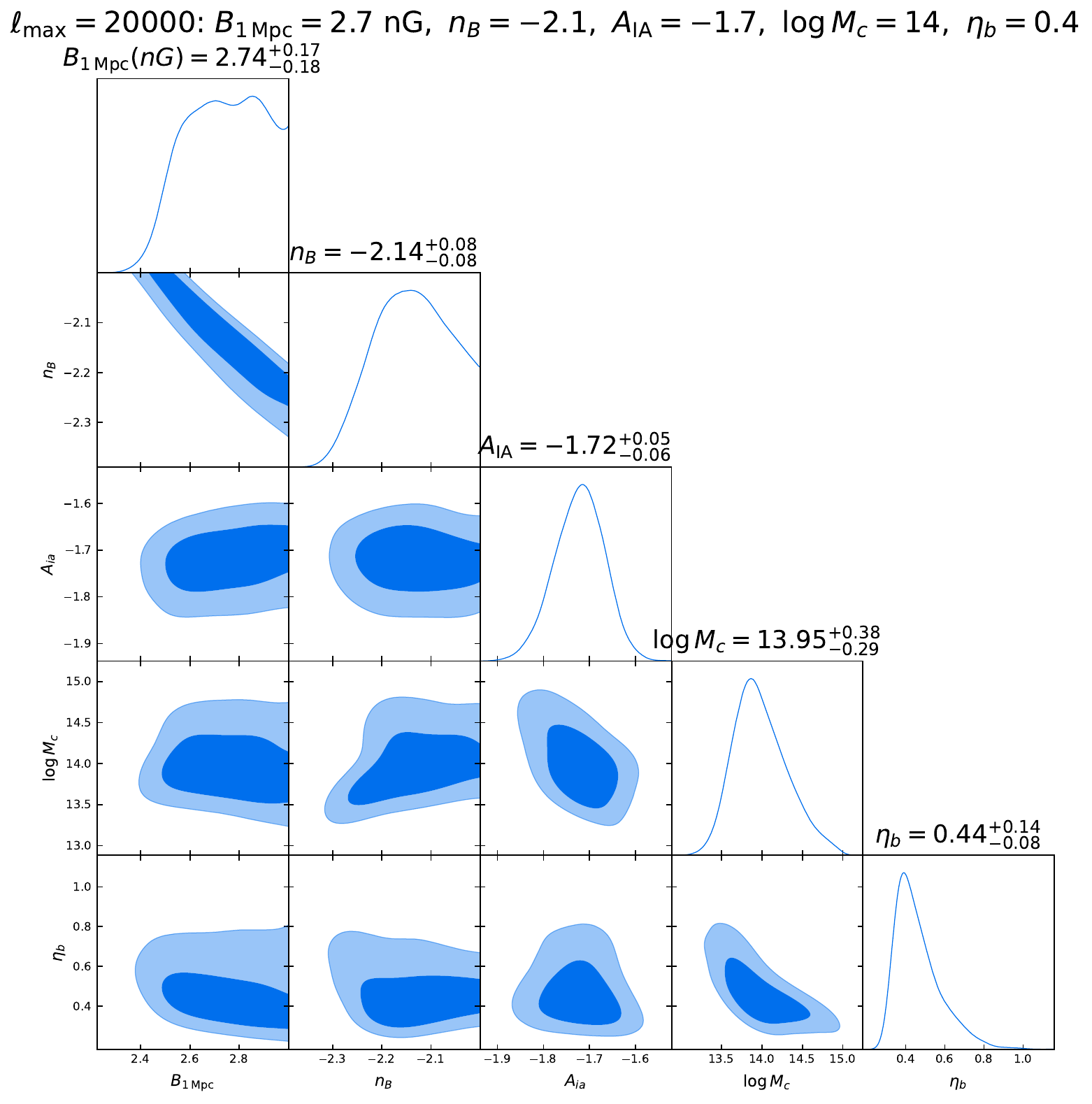}
\end{minipage}
\caption{Comparison of the posterior distributions. On the left the case $\ell_{\rm max}=5000$, $B_{\rm 1Mpc}=1.3$ nG and $n_B=-2.1$; on the right $\ell_{\rm max}=20000$, $B_{\rm 1Mpc}=2.7$ nG and $n_B=-2.1$. The constraints on $B_{\rm 1Mpc}$ are expressed in nG. In both cases, we can see no degeneracy between the PMFs parameters and the other parameters, and a degeneracy between the PMFs parameters themselves.}
\label{fig:contours_comparison}
\end{figure}

\begin{figure}
    \centering
    \includegraphics[width=1.0\linewidth]{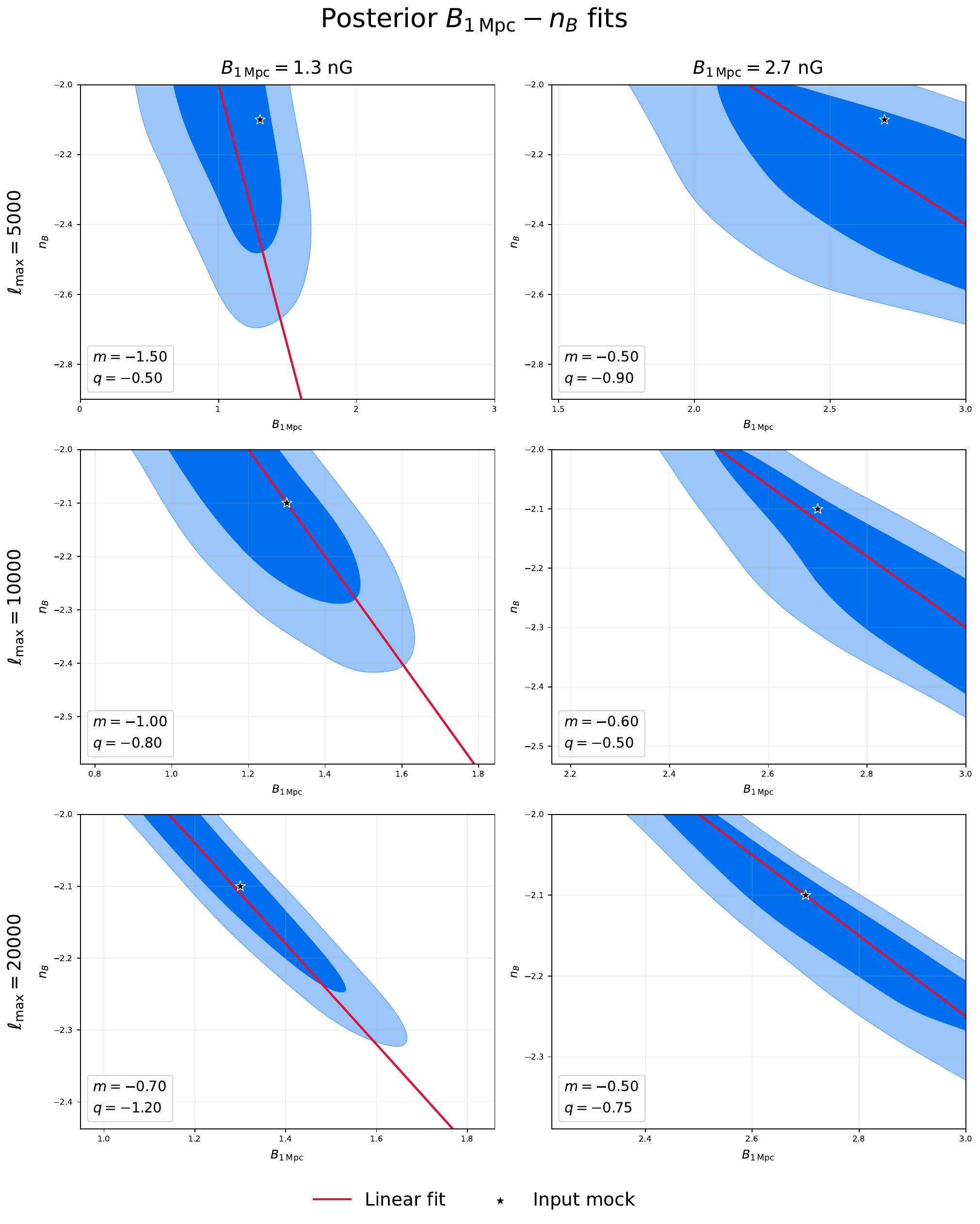}
    \caption{Two-dimensional posterior distributions of $B_{\rm 1Mpc}$-$n_B$ for the cases showing the clearest degeneracy. The red line represents the linear fit done by hand; the corresponding slopes and intercepts are reported in the lower-left corner of each panel.}
    \label{fig:linear fit}
\end{figure}

To create the collection of data mocks, we fix the cosmology to \textit{Planck} \cite{Planck:2015fie} and the other parameters to $A_{\rm IA}=-1.7$, $\log M_c=14$, $\eta_b=0.4$, $z_c=2.0$, keeping only $B_{\rm 1Mpc}$ and $n_B$ varying, with $B_{\rm 1Mpc} \in \{0.0, 0.3, 1.3, 2.7\}$ nG, and $n_B \in \{-2.7,-2.1\}$. In Figure \ref{fig:Ck data mocks}, at the top of each subplot we plot the data mocks and the $\Lambda$CDM convergence power spectrum for each redshift bin, and at the bottom the residuals of the PMFs results with respect to the $\Lambda$CDM one. We can see how the introduction of PMFs in the theory results in an increase of the convergence power spectrum at very large multipoles. As expected, for the higher values of $B_{\rm 1Mpc}$, for which the bump in the matter power spectrum is more prominent, the increase is larger.
 \\
In the MCMC runs we vary PMFs, baryon feedback and intrinsic alignment parameters in the range reported in Table \ref{tab:mcmc priors ranges}.
\begin{table}
\begin{center}
\begin{tabular}{|c|c|}
\hline
$B_{\rm 1Mpc}$ & $[0, 3.0] \text{nG}$ \\
\hline
$n_B$ & $[-2.9, -2.0]$ \\
\hline
$A_{\rm IA}$ & $[-2, 0]$ \\
\hline
$\log M_c$ & $[12,15]$ \\
\hline
$\eta_b$ & $[0, 2]$ \\
\hline
\end{tabular}
\end{center}
\caption{Priors used for the $5$ free parameters in the MCMC analysis.}
\label{tab:mcmc priors ranges}
\end{table}
We repeat the MCMC run and analysis for different values of $\ell_{\rm max}$, which is the maximum multipole of the $ 10$-point logarithmically spaced multipoles between $10$ and $\ell_{\rm max}$, used in the MCMC. We consider $\ell_{\rm max}=5000$, $10000$ and $20000$, to explore respectively a value that this generation of experiments could observe 
\cite{Euclid:2019clj}, a more extreme next generation scenario and a value far from what can currently be observed, but useful for fully showcasing our methodology's capability.
To run the MCMCs, we use \textit{emcee}, a Python implementation of the affine-invariant ensemble sampler for Markov chain Monte Carlo from \cite{10.2140/camcos.2010.5.65,foreman2013emcee}, in which the Metropolis-Hastings method is used. \\
In Table \ref{tab:mcmc results} we report the results for all MCMC runs, the eight data mocks for each of the three values of $\ell_{\rm max}$: for each parameter we write the median value and the $68\%$ confidence level, except when $B_{\rm 1Mpc}^{\rm mock}=0$ where we write the upper limit with the $95\%$ confidence level.
\begin{table}
\footnotesize
\renewcommand{\arraystretch}{1.35}
\makebox[\textwidth][c]{%
\begin{tabular}{c|ccccc|ccccc|c}
\cline{2-11}
                                             & \multicolumn{5}{c|}{Data mock}                                                                                                                 & \multicolumn{5}{c|}{MCMC results}                                                               & \multicolumn{1}{c}{}                                             \\ \cline{1-11}
\multicolumn{1}{|c|}{$\ell_{\rm max}$}       & \multicolumn{1}{c|}{$B_{\rm 1Mpc}$ (nG)} & \multicolumn{1}{c|}{$n_B$}  & \multicolumn{1}{c|}{$A_{\rm IA}$} & \multicolumn{1}{c|}{$\log M_C$}  & $\eta_b$ & \multicolumn{1}{c|}{$B_{\rm 1Mpc}$ (nG)} & \multicolumn{1}{c|}{$n_B$} & \multicolumn{1}{c|}{$A_{\rm IA}$} & \multicolumn{1}{c|}{$\log M_C$} & $\eta_b$ & \multicolumn{1}{c}{} \\ \Xhline{1.2pt}
\multicolumn{1}{|c|}{\multirow{8}{*}{5000}}  & \multicolumn{1}{c|}{$0.0$}          & \multicolumn{1}{c|}{$-2.1$} & \multicolumn{1}{c|}{$-1.7$}       & \multicolumn{1}{c|}{$14.0$} & $0.4$    & \multicolumn{1}{c|}{$<1.62$}            & \multicolumn{1}{c|}{$-2.45^{+0.29}_{-0.30}$}      & \multicolumn{1}{c|}{$-1.71^{+0.06}_{-0.06}$}             & \multicolumn{1}{c|}{$14.02^{+0.38}_{-0.34}$}      &   \multicolumn{1}{c|}{$0.39^{+0.10}_{-0.08}$}   &  \textcolor{YellowOrange}{\faCircleO}  \\ \cline{2-12} 
\multicolumn{1}{|c|}{}                       & \multicolumn{1}{c|}{$0.0$}          & \multicolumn{1}{c|}{$-2.7$} & \multicolumn{1}{c|}{$-1.7$}       & \multicolumn{1}{c|}{$14.0$} & $0.4$    & \multicolumn{1}{c|}{$<1.64$}               & \multicolumn{1}{c|}{$-2.45^{+0.30}_{-0.31}$}      & \multicolumn{1}{c|}{$-1.71^{+0.06}_{-0.06}$}             & \multicolumn{1}{c|}{$14.03^{+0.36}_{-0.35}$}      &    \multicolumn{1}{c|}{$0.39^{+0.10}_{-0.08}$}  &  \textcolor{YellowOrange}{\faCircleO}    \\ \cline{2-12} 
\multicolumn{1}{|c|}{}                       & \multicolumn{1}{c|}{$0.3$}          & \multicolumn{1}{c|}{$-2.1$} & \multicolumn{1}{c|}{$-1.7$}       & \multicolumn{1}{c|}{$14.0$} & $0.4$    & \multicolumn{1}{c|}{$0.81^{+0.57}_{-0.57}$}               & \multicolumn{1}{c|}{$-2.47^{+0.31}_{-0.30}$}      & \multicolumn{1}{c|}{$-1.70^{+0.06}_{-0.06}$}             & \multicolumn{1}{c|}{$14.03^{+0.40}_{-0.37}$}      &    \multicolumn{1}{c|}{$0.39^{+0.10}_{-0.08}$}  & \textcolor{Red}{\ding{55}}   \\ \cline{2-12} 
\multicolumn{1}{|c|}{}                       & \multicolumn{1}{c|}{$0.3$}          & \multicolumn{1}{c|}{$-2.7$} & \multicolumn{1}{c|}{$-1.7$}       & \multicolumn{1}{c|}{$14.0$} & $0.4$    & \multicolumn{1}{c|}{$0.84^{+0.53}_{-0.58}$}               & \multicolumn{1}{c|}{$-2.44^{+0.30}_{-0.31}$}      & \multicolumn{1}{c|}{$-1.71^{+0.06}_{-0.06}$}             & \multicolumn{1}{c|}{$14.02^{+0.41}_{-0.40}$}      &     \multicolumn{1}{|c|}{$0.39^{+0.11}_{-0.09}$} &\textcolor{Red}{\ding{55}}    \\ \cline{2-12} 
\multicolumn{1}{|c|}{}                       & \multicolumn{1}{c|}{$1.3$}          & \multicolumn{1}{c|}{$-2.1$} & \multicolumn{1}{c|}{$-1.7$}       & \multicolumn{1}{c|}{$14.0$} & $0.4$    & \multicolumn{1}{c|}{$1.15^{+0.22}_{-0.28}$}               & \multicolumn{1}{c|}{$-2.22^{+0.15}_{-0.21}$}      & \multicolumn{1}{c|}{$-1.73^{+0.06}_{-0.07}$}             & \multicolumn{1}{c|}{$14.14^{+0.31}_{-0.23}$}      &     \multicolumn{1}{|c|}{$0.46^{+0.08}_{-0.06}$}  & \textcolor{Green}{\ding{51}}   \\ \cline{2-12} 
\multicolumn{1}{|c|}{}                       & \multicolumn{1}{c|}{$1.3$}          & \multicolumn{1}{c|}{$-2.7$} & \multicolumn{1}{c|}{$-1.7$}       & \multicolumn{1}{c|}{$14.0$} & $0.4$    & \multicolumn{1}{c|}{$0.91^{+0.46}_{-0.62}$}               & \multicolumn{1}{c|}{$-2.41^{+0.28}_{-0.32}$}      & \multicolumn{1}{c|}{$-1.71^{+0.06}_{-0.06}$}             & \multicolumn{1}{c|}{$14.08^{+0.37}_{-0.36}$}      &     \multicolumn{1}{|c|}{$0.41^{+0.10}_{-0.08}$} & \textcolor{Red}{\ding{55}}   \\ \cline{2-12} 
\multicolumn{1}{|c|}{}                       & \multicolumn{1}{c|}{$2.7$}          & \multicolumn{1}{c|}{$-2.1$} & \multicolumn{1}{c|}{$-1.7$}       & \multicolumn{1}{c|}{$14.0$} & $0.4$    & \multicolumn{1}{c|}{$2.60^{+0.28}_{-0.39}$}               & \multicolumn{1}{c|}{$-2.27^{+0.17}_{-0.19}$}      & \multicolumn{1}{c|}{$-1.74^{+0.07}_{-0.07}$}             & \multicolumn{1}{c|}{$13.54^{+0.55}_{-0.29}$}      &     \multicolumn{1}{|c|}{$0.69^{+0.45}_{-0.30}$}    & \textcolor{Green}{\ding{51}} \\ \cline{2-12} 
\multicolumn{1}{|c|}{}                       & \multicolumn{1}{c|}{$2.7$}          & \multicolumn{1}{c|}{$-2.7$} & \multicolumn{1}{c|}{$-1.7$}       & \multicolumn{1}{c|}{$14.0$} & $0.4$    & \multicolumn{1}{c|}{$1.12^{+0.68}_{-0.71}$}               & \multicolumn{1}{c|}{$-2.46^{+0.32}_{-0.29}$}      & \multicolumn{1}{c|}{$-1.69^{+0.07}_{-0.06}$}             & \multicolumn{1}{c|}{$13.42^{+0.51}_{-0.36}$}      &  \multicolumn{1}{c|}{$0.69^{+0.37}_{-0.29}$}  & \textcolor{Red}{\ding{55}}   \\ \Xhline{1.2pt}
\multicolumn{1}{|c|}{\multirow{8}{*}{10000}} & \multicolumn{1}{c|}{$0.0$}          & \multicolumn{1}{c|}{$-2.1$} & \multicolumn{1}{c|}{$-1.7$}       & \multicolumn{1}{c|}{$14.0$} & $0.4$    & \multicolumn{1}{c|}{$<1.04$}               & \multicolumn{1}{c|}{$-2.60^{+0.40}_{-0.23}$}      & \multicolumn{1}{c|}{$-1.70^{+0.05}_{-0.05}$}             & \multicolumn{1}{c|}{$14.19^{+0.28}_{-0.25}$}      &     \multicolumn{1}{|c|}{$0.38^{+0.04}_{-0.04}$} &  \textcolor{YellowOrange}{\faCircleO}    \\ \cline{2-12} 
\multicolumn{1}{|c|}{}                       & \multicolumn{1}{c|}{$0.0$}          & \multicolumn{1}{c|}{$-2.7$} & \multicolumn{1}{c|}{$-1.7$}       & \multicolumn{1}{c|}{$14.0$} & $0.4$    & \multicolumn{1}{c|}{$<1.05$}               & \multicolumn{1}{c|}{$-2.59^{+0.40}_{-0.23}$}      & \multicolumn{1}{c|}{$-1.70^{+0.05}_{-0.06}$}             & \multicolumn{1}{c|}{$14.17^{+0.29}_{-0.23}$}      &     \multicolumn{1}{|c|}{$0.39^{+0.04}_{-0.04}$}  &  \textcolor{YellowOrange}{\faCircleO}   \\ \cline{2-12} 
\multicolumn{1}{|c|}{}                       & \multicolumn{1}{c|}{$0.3$}          & \multicolumn{1}{c|}{$-2.1$} & \multicolumn{1}{c|}{$-1.7$}       & \multicolumn{1}{c|}{$14.0$} & $0.4$    & \multicolumn{1}{c|}{$0.41^{+0.49}_{-0.27}$}               & \multicolumn{1}{c|}{$-2.58^{+0.37}_{-0.24}$}      & \multicolumn{1}{c|}{$-1.71^{+0.05}_{-0.06}$}             & \multicolumn{1}{c|}{$14.16^{+0.31}_{-0.29}$}      &     \multicolumn{1}{|c|}{$0.39^{+0.06}_{-0.06}$} &  \textcolor{YellowOrange}{\faCircleO}    \\ \cline{2-12} 
\multicolumn{1}{|c|}{}                       & \multicolumn{1}{c|}{$0.3$}          & \multicolumn{1}{c|}{$-2.7$} & \multicolumn{1}{c|}{$-1.7$}       & \multicolumn{1}{c|}{$14.0$} & $0.4$    & \multicolumn{1}{c|}{$0.36^{+0.49}_{-0.25}$}               & \multicolumn{1}{c|}{$-2.58^{+0.37}_{-0.24}$}      & \multicolumn{1}{c|}{$-1.71^{+0.05}_{-0.06}$}             & \multicolumn{1}{c|}{$14.15^{+0.32}_{-0.27}$}      &     \multicolumn{1}{|c|}{$0.39^{+0.05}_{-0.05}$}  &  \textcolor{YellowOrange}{\faCircleO}   \\ \cline{2-12} 
\multicolumn{1}{|c|}{}                       & \multicolumn{1}{c|}{$1.3$}          & \multicolumn{1}{c|}{$-2.1$} & \multicolumn{1}{c|}{$-1.7$}       & \multicolumn{1}{c|}{$14.0$} & $0.4$    & \multicolumn{1}{c|}{$1.28^{+0.16}_{-0.15}$}               & \multicolumn{1}{c|}{$-2.14^{+0.10}_{-0.13}$}      & \multicolumn{1}{c|}{$-1.73^{+0.05}_{-0.06}$}             & \multicolumn{1}{c|}{$14.01^{+0.28}_{-0.20}$}      &     \multicolumn{1}{|c|}{$0.46^{+0.10}_{-0.06}$}     & \textcolor{Green}{\ding{51}} \\ \cline{2-12} 
\multicolumn{1}{|c|}{}                       & \multicolumn{1}{c|}{$1.3$}          & \multicolumn{1}{c|}{$-2.7$} & \multicolumn{1}{c|}{$-1.7$}       & \multicolumn{1}{c|}{$14.0$} & $0.4$    & \multicolumn{1}{c|}{$0.77^{+0.42}_{-0.42}$}               & \multicolumn{1}{c|}{$-2.52^{+0.33}_{-0.24}$}      & \multicolumn{1}{c|}{$-1.73^{+0.06}_{-0.06}$}             & \multicolumn{1}{c|}{$14.08^{+0.33}_{-0.24}$}      &     \multicolumn{1}{|c|}{$0.44^{+0.09}_{-0.08}$}  & \textcolor{Red}{\ding{55}}   \\ \cline{2-12} 
\multicolumn{1}{|c|}{}                       & \multicolumn{1}{c|}{$2.7$}          & \multicolumn{1}{c|}{$-2.1$} & \multicolumn{1}{c|}{$-1.7$}       & \multicolumn{1}{c|}{$14.0$} & $0.4$    & \multicolumn{1}{c|}{$2.79^{+0.15}_{-0.18}$}               & \multicolumn{1}{c|}{$-2.20^{+0.11}_{-0.11}$}      & \multicolumn{1}{c|}{$-1.72^{+0.06}_{-0.06}$}             & \multicolumn{1}{c|}{$13.76^{+0.52}_{-0.33}$}      &     \multicolumn{1}{|c|}{$0.50^{+0.25}_{-0.15}$}     & \textcolor{Green}{\ding{51}} \\ \cline{2-12} 
\multicolumn{1}{|c|}{}                       & \multicolumn{1}{c|}{$2.7$}          & \multicolumn{1}{c|}{$-2.7$} & \multicolumn{1}{c|}{$-1.7$}       & \multicolumn{1}{c|}{$14.0$} & $0.4$    & \multicolumn{1}{c|}{$1.32^{+0.78}_{-0.51}$}               & \multicolumn{1}{c|}{$-2.53^{+0.28}_{-0.18}$}      & \multicolumn{1}{c|}{$-1.70^{+0.06}_{-0.06}$}             & \multicolumn{1}{c|}{$13.28^{+0.64}_{-0.30}$}      &      \multicolumn{1}{c|}{$0.67^{+0.43}_{-0.30}$} & \textcolor{Red}{\ding{55}}   \\ \Xhline{1.2pt}
\multicolumn{1}{|c|}{\multirow{8}{*}{20000}} & \multicolumn{1}{c|}{$0.0$}          & \multicolumn{1}{c|}{$-2.1$} & \multicolumn{1}{c|}{$-1.7$}       & \multicolumn{1}{c|}{$14.0$} & $0.4$    & \multicolumn{1}{c|}{$<0.80$}               & \multicolumn{1}{c|}{$-2.62^{+0.35}_{-0.22}$}      & \multicolumn{1}{c|}{$-1.71^{+0.04}_{-0.04}$}             & \multicolumn{1}{c|}{$14.16^{+0.25}_{-0.23}$}      &       \multicolumn{1}{|c|}{$0.40^{+0.03}_{-0.03}$} &  \textcolor{YellowOrange}{\faCircleO}  \\ \cline{2-12} 
\multicolumn{1}{|c|}{}                       & \multicolumn{1}{c|}{$0.0$}          & \multicolumn{1}{c|}{$-2.7$} & \multicolumn{1}{c|}{$-1.7$}       & \multicolumn{1}{c|}{$14.0$} & $0.4$    & \multicolumn{1}{c|}{$<0.78$}               & \multicolumn{1}{c|}{$-2.61^{+0.34}_{-0.23}$}      & \multicolumn{1}{c|}{$-1.71^{+0.04}_{-0.04}$}             & \multicolumn{1}{c|}{$14.16^{+0.25}_{-0.22}$}      &       \multicolumn{1}{|c|}{$0.40^{+0.03}_{-0.03}$} &  \textcolor{YellowOrange}{\faCircleO}  \\ \cline{2-12} 
\multicolumn{1}{|c|}{}                       & \multicolumn{1}{c|}{$0.3$}          & \multicolumn{1}{c|}{$-2.1$} & \multicolumn{1}{c|}{$-1.7$}       & \multicolumn{1}{c|}{$14.0$} & $0.4$    & \multicolumn{1}{c|}{$0.41^{+0.47}_{-0.26}$}               & \multicolumn{1}{c|}{$-2.59^{+0.41}_{-0.23}$}      & \multicolumn{1}{c|}{$-1.72^{+0.05}_{-0.05}$}             & \multicolumn{1}{c|}{$14.07^{+0.30}_{-0.24}$}      &       \multicolumn{1}{|c|}{$0.42^{+0.06}_{-0.05}$}  &  \textcolor{YellowOrange}{\faCircleO}  \\ \cline{2-12} 
\multicolumn{1}{|c|}{}                       & \multicolumn{1}{c|}{$0.3$}          & \multicolumn{1}{c|}{$-2.7$} & \multicolumn{1}{c|}{$-1.7$}       & \multicolumn{1}{c|}{$14.0$} & $0.4$    & \multicolumn{1}{c|}{$0.25^{+0.36}_{-0.18}$}               & \multicolumn{1}{c|}{$-2.62^{+0.36}_{-0.22}$}      & \multicolumn{1}{c|}{$-1.72^{+0.05}_{-0.05}$}             & \multicolumn{1}{c|}{$14.13^{+0.29}_{-0.27}$}      &       \multicolumn{1}{|c|}{$0.40^{+0.04}_{-0.04}$} &   \textcolor{Green}{\ding{51}}  \\ \cline{2-12} 
\multicolumn{1}{|c|}{}                       & \multicolumn{1}{c|}{$1.3$}          & \multicolumn{1}{c|}{$-2.1$} & \multicolumn{1}{c|}{$-1.7$}       & \multicolumn{1}{c|}{$14.0$} & $0.4$    & \multicolumn{1}{c|}{$1.32^{+0.16}_{-0.12}$}               & \multicolumn{1}{c|}{$-2.12^{+0.08}_{-0.10}$}      & \multicolumn{1}{c|}{$-1.72^{+0.05}_{-0.05}$}             & \multicolumn{1}{c|}{$14.05^{+0.25}_{-0.19}$}      &       \multicolumn{1}{|c|}{$0.43^{+0.06}_{-0.05}$}    & \textcolor{Green}{\ding{51}} \\ \cline{2-12} 
\multicolumn{1}{|c|}{}                       & \multicolumn{1}{c|}{$1.3$}          & \multicolumn{1}{c|}{$-2.7$} & \multicolumn{1}{c|}{$-1.7$}       & \multicolumn{1}{c|}{$14.0$} & $0.4$    & \multicolumn{1}{c|}{$0.94^{+0.31}_{-0.32}$}               & \multicolumn{1}{c|}{$-2.49^{+0.28}_{-0.20}$}      & \multicolumn{1}{c|}{$-1.71^{+0.05}_{-0.05}$}             & \multicolumn{1}{c|}{$14.05^{+0.34}_{-0.24}$}      &    \multicolumn{1}{|c|}{$0.41^{+0.07}_{-0.05}$}    & \textcolor{Red}{\ding{55}}  \\ \cline{2-12} 
\multicolumn{1}{|c|}{}                       & \multicolumn{1}{c|}{$2.7$}          & \multicolumn{1}{c|}{$-2.1$} & \multicolumn{1}{c|}{$-1.7$}       & \multicolumn{1}{c|}{$14.0$} & $0.4$    & \multicolumn{1}{c|}{$2.74^{+0.17}_{-0.18}$}               & \multicolumn{1}{c|}{$-2.14^{+0.08}_{-0.08}$}      & \multicolumn{1}{c|}{$-1.72^{+0.05}_{-0.06}$}             & \multicolumn{1}{c|}{$13.95^{+0.38}_{-0.29}$}      &       \multicolumn{1}{|c|}{$0.44^{+0.14}_{-0.08}$}   & \textcolor{Green}{\ding{51}} \\ \cline{2-12} 
\multicolumn{1}{|c|}{}                       & \multicolumn{1}{c|}{$2.7$}          & \multicolumn{1}{c|}{$-2.7$} & \multicolumn{1}{c|}{$-1.7$}       & \multicolumn{1}{c|}{$14.0$} & $0.4$    & \multicolumn{1}{c|}{$1.81^{+0.78}_{-0.71}$}               & \multicolumn{1}{c|}{$-2.52^{+0.30}_{-0.16}$}      & \multicolumn{1}{c|}{$-1.69^{+0.05}_{-0.05}$}             & \multicolumn{1}{c|}{$13.74^{+0.49}_{-0.64}$} &     \multicolumn{1}{c|}{$0.42^{+0.17}_{-0.09}$} &    \textcolor{Red}{\ding{55}}    \\ \hline
\end{tabular}%
}
\caption{Results of the MCMCs for every combination of the maximum multipole considered ($\ell_{\rm max}$) and the parameters used to create the data mocks. In every case, we were able to retrieve the intrinsic alignment ($A_{IA}$) and baryon feedback ($\log M_c, \eta_b$) parameters. For the case $B_{\rm 1Mpc}^{\rm mock}=0$, we report the upper limit at $95\%$ confidence level. On the last column on the right, we indicated with a different symbol if we were able to recover only $B_{\rm 1Mpc}$ (\textcolor{YellowOrange}{\faCircleO}), both $B_{\rm 1Mpc}$ and $n_B$ (\textcolor{Green}{\ding{51}}), or neither of them (\textcolor{Red}{\ding{55}}).}
\label{tab:mcmc results}
\end{table}
The marginalized posteriors of the intrinsic-alignment and baryon-feedback parameters are consistent with their input values in all runs. We classify a parameter as recovered when its input value lies within the marginalized $68\%$ credible interval and the posterior is sufficiently constrained relative to the adopted prior. Looking at the PMFs parameters, for every case we have one of three scenarios: we recover both input values; we recover $B_{\rm 1Mpc}$ within the adopted criterion while $n_B$ remains unconstrained; neither of the two parameters can be recovered. In the last column on the right of Table \ref{tab:mcmc results} we indicate the scenario for each data mock using a green tick (\textcolor{Green}{\ding{51}}) for the first scenario, an orange circle (\textcolor{YellowOrange}{\faCircleO}) for the second and a red cross (\textcolor{Red}{\ding{55}}) for the third. For the $\Lambda$CDM mocks, corresponding to $B_{\rm 1Mpc}^{\rm mock}=0$, we obtain $95\%$ upper limits of approximately $B_{\rm 1Mpc}<1.6\,{\rm nG}$, $1.0\,{\rm nG}$, and $0.8\,{\rm nG}$ for $\ell_{\rm max}=5000$, $10000$, and $20000$, respectively. The limits depend only weakly on the input value of $n_B$. Our results can be compared with \cite{Zucca:2016iur}, who used \textit{Planck}+SPT data: they obtained $B_{\rm 1Mpc}<1.5$ nG at the $95\%$ confidence level (CL) when they marginalized over $n_B$, and $B_{\rm 1Mpc}<1.2$ nG at the $95\%$ CL when they considered $n_B=-2.9$. Other recent works have identified upper limits on $B_{\rm 1Mpc}$ at the $95\%$ confidence level, but with $n_B$ fixed: $B_{\rm 1Mpc}<0.3$ nG with $n_B=-2.9$, using the Lyman-$\alpha$ forest \cite{Pavicevic:2025gqi}; $B_{\rm 1Mpc}<0.95$ nG with $n_B=-2.9$ and $B_{\rm 1Mpc}<0.22$ nG with $n_B=-2$, using \textit{Planck2018} data \cite{Paoletti:2022gsn}; $B_{\rm 1Mpc}<0.04$ nG with $n_B=-2$, using JWST data \cite{2026arXiv260424835F}. When $B_{\rm 1Mpc}=0$, the observable is independent of $n_B$, so the spectral index is not physically identifiable. The marginalized values of $n_B$ in these runs should not be interpreted as valid constraints. For all values of $\ell_{\rm max}$, when $n_B=-2.7$, we are never able to get recover $n_B$, except for $\ell_{\rm max}=20000$ and $B_{\rm 1Mpc}=0.3$ nG, and only in a few cases we can recover $B_{\rm 1Mpc}$; this is because a more negative value of $n_B$ corresponds to a less pronounced bump in $P_m(k)$, so a less visible effect in $C_{\kappa}(\ell)$. When $n_B=-2.1$, we can always recover $B_{\rm 1Mpc}$, except for $B_{\rm 1Mpc}=0.3$ nG in the $\ell_{\rm max}=5000$ case, where the deviation from $\Lambda$CDM is too small to make the effect of PMFs visible. The constraining power on $B_{\rm 1Mpc}$ improves with larger $\ell_{\rm max}$: already for $\ell_{\rm max}=10000$ constraints on $B_{\rm 1Mpc}$ are consistent with the input value even for $B_{\rm 1Mpc}=0.3$ nG. \\
For $\ell_{\rm max}=5000$, in the case $B_{\rm 1Mpc}^{\rm mock}=1.3$ nG and $n_B^{\rm mock}=-2.1$, the $3\sigma$ confidence level calculated on the $B_{\rm 1Mpc}$ sample is $1.15^{+0.53}_{-1.10}$ nG, which would make it possible to distinguish between this model and the $\Lambda$CDM model. The situation improves if we consider $\ell_{\rm max}=10000$, in which case we have $1.28^{+0.42}_{-0.39}$ nG, and for $B_{\rm 1Mpc}^{\rm mock}=0.3$ nG the $3\sigma$ confidence level is $0.41^{+2.17}_{-0.40}$ nG, so it is still identifiable.\\
In Figure \ref{fig:contours_comparison} we show the $1\sigma$ and $2\sigma$ contours plots for two cases: for $\ell_{\rm max}=5000$, $B_{\rm 1Mpc}=1.3$ nG, $n_B=-2.1$ (Left), and for $\ell_{\rm max}=20000$, $B_{\rm 1Mpc}=2.7$ nG, $n_B=-2.1$ (Right).
We can observe how there is no degeneracy between the PMFs and the baryon feedback and intrinsic alignment parameters, while for some models there is a degeneracy between $B_{\rm 1Mpc}$ and $n_B$. In particular, we perform a linear fit by hand with the formula
\begin{align}\label{eq:Linear fit}
    n_B=m \cdot B_{\rm 1Mpc} + q
\end{align}
of the $2$D $B_{\rm 1Mpc}$-$n_B$ posterior distribution for the cases where the degeneracy was more evident. We considered $n_B=-2.1$, $B_{\rm 1Mpc}=1.3 {\rm \ nG}, \ 2.7 {\rm \ nG}$. In Table \ref{tab:linear fit} we report the results of the fit and in Figure \ref{fig:linear fit} we show the fit on top of the $2$D posterior distribution for each case considered.
\begin{table}
\begin{center}
\begin{tabular}{|c|c|c|c|c|}
\hline
\multicolumn{3}{|c|}{Mock parameters} & \multicolumn{2}{c|}{Linear fit results} \\
\hline
$\ell_{\rm max}$ & $B_{\rm 1Mpc}$(nG) & $n_B$ & $m$ & $q$ \\
\hline
$5000$ & $1.3$ & $-2.1$ & $-1.50$ & $-0.50$\\
\hline
$5000$ & $2.7$ & $-2.1$ & $-0.50$ & $-0.90$\\
\hline
$10000$ & $1.3$ & $-2.1$ & $-1.00$ & $-0.80$\\
\hline
$10000$ & $2.7$ & $-2.1$ & $-0.60$ & $-0.50$\\
\hline
$20000$ & $1.3$ & $-2.1$ & $-0.70$ & $-1.20$\\
\hline
$20000$ & $2.7$ & $-2.1$ & $-0.50$ & $-0.75$\\
\hline
\end{tabular}
\end{center}
\caption{Results of the linear fit ($n_B=m \cdot B_{\rm 1Mpc} + q$) of the $B_{\rm 1Mpc}$-$n_B$ $2$D posterior distribution for the cases where there was a clear degeneracy between the two PMFs parameters.}
\label{tab:linear fit}
\end{table}

\section{Conclusions}\label{sec:conclusions}
In this study, we construct an emulator of the convergence power spectrum of weak lensing to forecast the constraints on the parameters characterizing the primordial magnetic fields power spectrum: its strength $B_{\rm 1Mpc}$ and power spectral index $n_B$. \\
We first solve the coupled ordinary differential equations that describe the evolution of baryon and dark matter perturbations in the case where we can neglect turbulence \cite{Ralegankar:2024dpb}. Assuming a power-law magnetic-field power spectrum, $P_B(k)\propto B_{\rm 1Mpc}^2 k^{n_B}$, we calculate the linear matter power spectrum for different combinations of PMFs parameters: $B_{\rm 1Mpc} \in \{0.25, 0.5, 1.0, 1.5, 2.0, 3.0\}$ nG $\times$ $n_B \in \{-2.9, -2.5, -2.0\}$. Adding the matter power spectrum due to PMFs to the one due to primordial fluctuations described by the $\Lambda$CDM model, which corresponds to $B_{\rm 1Mpc}=0$, we calculate initial conditions for $18$ dark-matter-only N-body simulations at $z=299$, plus the $\Lambda$CDM one, from which we extract the non-linear matter power spectrum at different redshifts, $z \in \{0, 0.3, 1, 2, 3, 4, 5, 9\}$. \\
We calculate the tomographic convergence power spectrum of weak lensing taking into consideration intrinsic alignment and baryon corrections through a baryon correction model \cite{Parimbelli_2019}, and we create an emulator of $C_{\kappa}(\ell;[B_{\rm 1Mpc}, n_B, A_{IA}, \log M_c, \eta_b])$. Introducing PMFs in $C_{\kappa}$ results in an increase at very large multipoles. The amplitude and scale dependence of this enhancement depend on both $B_{\rm 1Mpc}$ and $n_B$: larger magnetic-field amplitudes and less negative spectral indices generally produce a stronger signal that extends to lower multipoles.\\
To test the constraining power of this observable, using our emulator we perform MCMC analyses with data mocks of $C_{\kappa}(\ell)$ generated for different values of the PMFs, IA and BF parameters, fixing the cosmology to \textit{Planck}: $B_{\rm 1Mpc} \in \{0, 0.3, 1.3, 2.7\}$ nG, $n_B \in \{-2.7, -2.1\}$, $A_{IA}=-1.7$, $\log M_c=14$, $\eta_b=0.4$. We fix $z_c=2$ because we are not able to put any constraint on it, leaving $5$ free parameters. We consider a \textit{Euclid}-like setup, taking into account both the statistical and systematic errors affecting the theoretical model \cite{Parimbelli_2019}. We repeat our analysis for different values of $\ell_{\rm max}$, the maximum multipole of the mock data vector, to see if our treatment could already be applied to present experiments. The intrinsic alignment and baryon feedback parameters are consistent with their input values in all the runs, and we find no clear evidence for a strong degeneracy between these parameters and the PMF parameters. \\
For the $\Lambda$CDM mocks, we obtain approximate upper limits of $B_{\rm 1Mpc}<1.6$, $1.0$, and $0.8\,{\rm nG}$ at the $95\%$ confidence level for $\ell_{\rm max}=5000$, $10000$, and $20000$, respectively. As expected, including higher multipoles improves the upper limit because the PMF contribution is concentrated on small angular scales. These values should be compared to recent results ranging from $B_{\rm 1Mpc}<0.04$ nG to $B_{\rm 1Mpc}<1.5$ nG, obtained with JWST \cite{2026arXiv260424835F} and \textit{Planck}+SPT data \cite{Zucca:2016iur}, respectively, at a fixed power spectral index value of $-2$ for the former and marginalizing over $n_B$ for the latter. For mocks with a non-zero magnetic field amplitude, the recovery of the PMF parameters depends strongly on the input spectral index. Models with $n_B=-2.1$ are more easily constrained because they produce a stronger enhancement of the matter and convergence power spectra. In these cases, $B_{\rm 1Mpc}$ is recovered for most of the models considered, including smaller input values when higher multipoles are included. For $n_B=-2.7$, the PMF effect is weaker and the spectral index generally remains unconstrained. Increasing $\ell_{\rm max}$ improves the constraints on $B_{\rm 1Mpc}$, but does not guarantee an improvement in the constraints on $n_B$. \\
Looking at the $2$D $B_{\rm 1Mpc}$-$n_B$ posterior distributions, we notice a degeneracy in some cases between the two PMFs parameters. We characterize the orientation of this degeneracy with linear relations of the form $n_B=m B_{\rm 1Mpc}+q$ and we do not find a simple dependence on either $\ell_{\rm max}$ or the input value of $B_{\rm 1Mpc}$. \\
Our results show that small-scale weak lensing measurements can provide complementary constraints on primordial magnetic fields. 
Overall, we find that the non-linear evolution of matter power leaves a characteristic PMFs-induced excess in the small-scale matter distribution that propagates into cosmic shear, and its shape is sufficiently different from the adopted baryon feedback and intrinsic alignment effects that useful PMF information can survive marginalization over those astrophysical uncertainties.
We caution that the constraints presented here are forecasts based on synthetic data, a fixed background cosmology, a specific model of baryon feedback and intrinsic alignment, and theoretical predictions extending to very high multipoles. A direct application to observational data will require a more complete treatment of cosmological parameter degeneracies, observational systematics, baryonic uncertainties, and the theoretical accuracy of the non-linear matter power spectrum. \\
Within these assumptions, our analysis indicates that the convergence power spectrum is a promising observable for quantitatively probing the small-scale effects of primordial magnetic fields and it is highly complementary in terms of scales, redshifts and physical processes with respect to low-redshift measurements of magnetized cosmic filaments \cite{carrettivazza25}, the high-z Lyman-$\alpha$ forest bounds \cite{Pavicevic:2025gqi} and the recombination/CMB window \cite{jedamzik26,Paoletti:2022gsn}.

\acknowledgments
The authors are supported by the INFN INDARK grant. Simulations were partly performed on the Ulysses supercomputer at SISSA and partly on DiRAC CSD3 facility under project dp002. Postprocessing of the simulations was done on Ulysses.
MV is also supported by SISSA IDEAS grant, the INAF Theory Grant "Cosmological Investigation of the Cosmic Web", and by the Fondazione ICSC, Spoke 3 Astrophysics and Cosmos Observations, National Recovery and Resilience Plan Project ID CN\_00000013 ``Italian Research Center on High-Performance Computing, Big Data and Quantum Computing'' funded by MUR Missione 4 Componente 2 Investimento 1.4: Potenziamento strutture di ricerca e creazione di "campioni nazionali di R\&S (M4C2-19 )" - Next Generation EU (NGEU).
MV also acknowledges the support by INAF Theory grant `Cosmological Investigation of the Cosmic Web' (C93C23006820005). \\
Part of the simulations used in this work were performed using the Joliot Curie supercomputer at the Très Grand Centre de Calcul (TGCC) and the Cambridge Service for Data Driven Discovery (CSD3), part of which is operated by the University of Cambridge Research Computing on behalf of the STFC DiRAC HPC Facility (www.dirac.ac.uk). We acknowledge the Partnership for Advanced Computing in Europe (PRACE) for awarding us time on Joliot Curie in the 16th call. The DiRAC component of CSD3 was funded by BEIS capital funding via STFC capital grants ST/P002307/1 and ST/R002452/1 and STFC operations grant ST/R00689X/1. This work also used the DiRAC@Durham facility managed by the Institute for Computational Cosmology on behalf of the STFC DiRAC HPC Facility. The equipment was funded by BEIS capital funding via STFC capital grants ST/P002293/1 and ST/R002371/1, Durham University and STFC operations grant ST/R000832/1. DiRAC is part of the National e-Infrastructure. \\
AD thanks Edoardo Martinelli  for helpful conversations.

\appendix
\section{Check of the accuracy of ICs}\label{appendix:IC}
\begin{figure}
    \centering
    \includegraphics[width=0.88\textwidth]{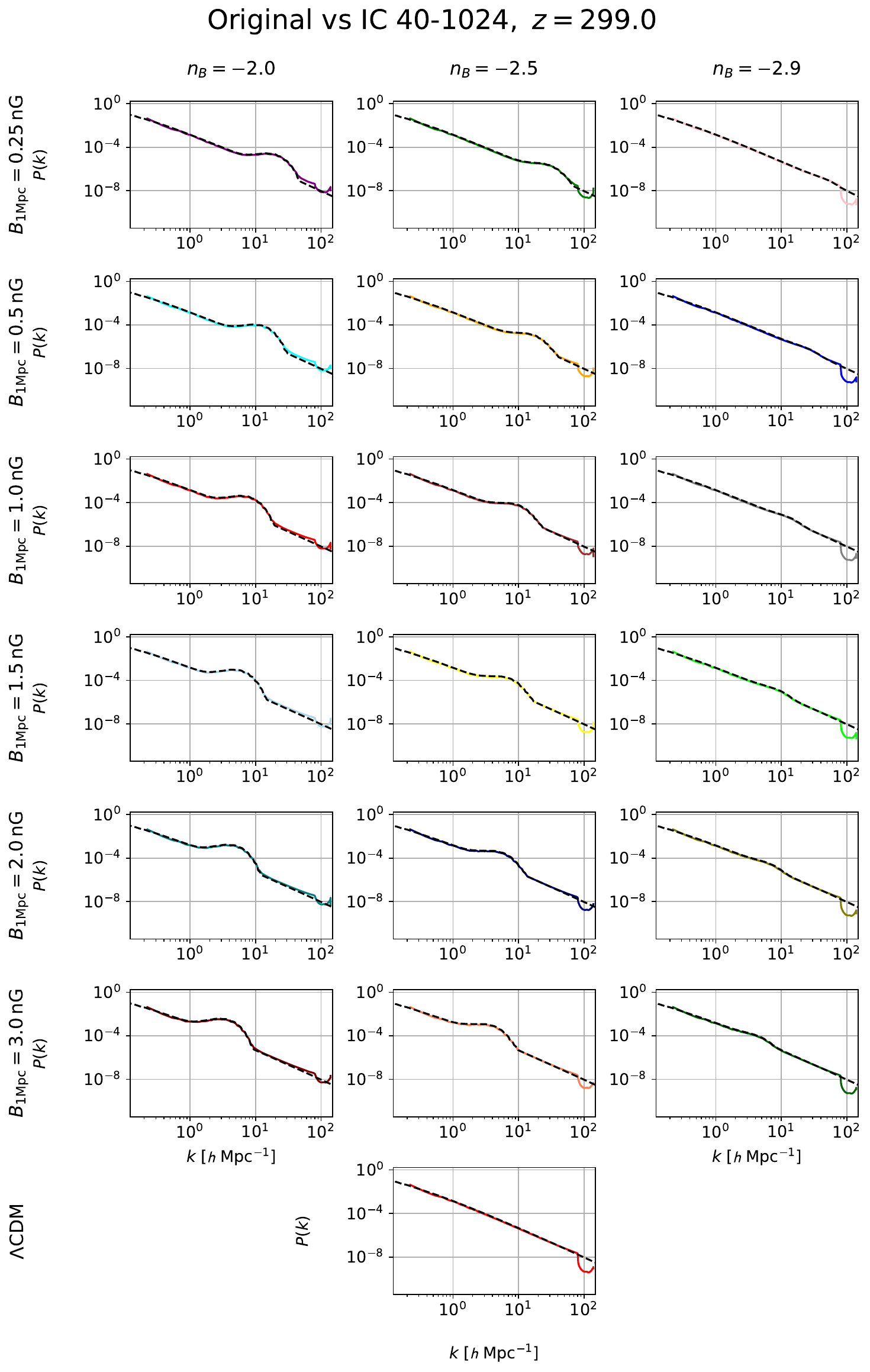}
    \caption{Initial conditions calculated with N-GenIC for different PMFs parameters and for $\Lambda$CDM (colored solid lines) compared to the original power spectrum (dashed black line) used as input to N-GenIC. The bump is well traced in ICs, while the valley at the end is above the Nyquist frequency, so it will not come into play in the analysis.}
    \label{fig:IC}
\end{figure}

Figure \ref{fig:IC} shows the initial conditions calculated with N-GenIC. The parameters and the value of $z$ are chosen so that the power spectra in the presence of PMFs come down on the original power spectra after the bump. The bump is traced well after the Zeldovich approximation is applied; the valley at the end is present because we are plotting values of $k$ above the Nyquist mode $\left( k_{Nyq} = \pi \frac{1024}{40} \approx 80 \right) h/\text{Mpc}$, so it does not pose a problem in the analysis. \\
\bibliographystyle{unsrt}
\bibliography{references}
	
\end{document}